\documentclass[aps,nofootinbib,prd,preprint,tightenlines,amsmath,amssymb]{revtex4-2}

\usepackage{bm}
\usepackage[colorlinks,anchorcolor=black,citecolor=violet,linkcolor=blue]{hyperref}
\usepackage{graphicx}
\usepackage{multirow,slashed}
\usepackage{xcolor}

\begin{document}
\baselineskip=17pt \parskip=5pt

\preprint{NCTS-PH/2011}

\title{Probing new physics with the kaon decays \boldmath$K\to\pi\pi\slashed E$}

\author{Chao-Qiang Geng$^{1,2,3,4}$ and Jusak Tandean$^{4,5}$ \bigskip \\ \it
$^1$School of Fundamental Physics and Mathematical Sciences, \\
Hangzhou Institute for Advanced Study, UCAS, Hangzhou 310024, China \medskip \\
$^2$International Centre for Theoretical Physics Asia-Pacific, Beijing/Hangzhou, China \medskip \\
$^3$Department of Physics, National Tsing Hua University, Hsinchu 300, Taiwan \medskip \\
$^4$Physics Division, National Center for Theoretical Sciences, Hsinchu 300, Taiwan \medskip \\
$^5$Department of Physics, National Taiwan University, Taipei 106, Taiwan \bigskip \\
\large\rm Abstract \medskip \\
\begin{minipage}{\textwidth} \baselineskip=17pt \parindent=3ex \small
The latest search for the rare kaon decay $K^+\to\pi^+\nu\bar\nu$ by the NA62 experiment
has produced  evidence for it with a branching fraction consistent with the prediction of
the standard model.
The new result implies that in this decay, with the $\nu\bar\nu$ pair appearing as missing
energy ($\slashed E$), the room for possible new physics is no longer sizable and that therefore
its contributions to underlying four-particle $s\to d\slashed E$ operators with parity-even $ds$ quark
bilinears have become significantly constrained.
Nevertheless, we point out that appreciable manifestations from beyond the standard model induced
by the corresponding operators with mainly parity-odd $ds$ quark bilinears could still occur in
$K\to\pi\pi\slashed E$ modes, on which there are only minimal empirical details at present.
We find in particular that new physics of this kind may enhance the branching fraction of
$K_L\to\pi^0\pi^0\slashed E$ to values reaching its current experimental upper limit and
the branching fractions of $K^+\to\pi^+\pi^0\slashed E$ and $K_L\to\pi^+\pi^-\slashed E$
to the levels of $10^{-7}$ and $10^{-6}$, respectively.
Thus, quests for these decays in existing kaon facilities such as KOTO and NA62 or future ones
could provide valuable information complementary to that gained from $K\to\pi\slashed E$.
\end{minipage}}


\maketitle

\section{Introduction\label{intro}}

One of the potentially promising avenues to discover new physics (NP) beyond the standard model (SM)
is to look for processes that are expected to be very rare in the SM.
An observation of such a process having a rate much greater than what the SM predicts would then
be a compelling indication of NP effects.
Among places where this may be realized are the flavor-changing neutral current (FCNC) decays of
light strange-flavored hadrons with missing energy ($\slashed E$).
These reactions are known to be dominated by short-distance physics~\cite{Littenberg:1993qv,
Buchalla:1995vs,Geng:1994cw,Geng:1996kd,Cirigliano:2011ny,Tandean:2019tkm,Su:2019tjn,Li:2019cbk}
and arise primarily from the quark transition \,$s\to d\slashed E$.
In the SM, it proceeds from loop-suppressed diagrams~\cite{Buchalla:1995vs} and the final state
contains undetected neutrinos ($\nu\bar\nu$).
Beyond the SM, there could be additional ingredients which alter the SM component and/or give
rise to extra channels with one or more invisible nonstandard particles carrying away
the missing energy.

Over the years hunts for \,$s\to d\slashed E$\, have focused the kaon modes
\,$K\to\pi\nu\bar\nu$,\, leading mostly to limits on their branching
fractions~\cite{Artamonov:2008qb,Ahn:2018mvc,CortinaGil:2018fkc,CortinaGil:2020vlo}.
The efforts are ongoing in the KOTO~\cite{Ahn:2018mvc} and
NA62~\cite{CortinaGil:2020vlo} experiments.
The former~\cite{Ahn:2018mvc} has set
${\mathcal B}(K_L\to\pi^0\nu\bar\nu)_{\textsc{koto}}<3.0\times10^{-9}$ at 90\% confidence level
(CL), exceeding but not far from the SM expectation~\cite{Buras:2015qea} of
\,${\mathcal B}(K_L\to\pi^0\nu\bar\nu)_{\textsc{sm}}=(3.4\pm0.6)\times 10^{-11}$.\,
On the other hand, very recently NA62~\cite{na62new} has preliminarily reported 3.5$\sigma$
evidence for the charged channel with
\,${\cal B}(K^+\to\pi^+\nu\bar\nu)_{\textsc{na}\scriptscriptstyle62} =
\big[11.0_{-3.5}^{+4.0}{\rm(stat)}\pm0.3{\rm(syst)}\big]\times 10^{-11}$,\,
which is in good agreement with the SM value~\cite{Buras:2015qea} of
\,${\cal B}(K^+\to\pi^+\nu\bar\nu)_{\textsc{sm}}=(8.4\pm1.0)\times 10^{-11}$\,
and more precise than the earlier E949~\cite{Artamonov:2008qb} finding of
\,${\cal B}(K^+\to\pi^+\nu\bar\nu)_{\textsc e\scriptscriptstyle949} =
\big(17.3_{-10.5}^{+11.5}\big)\times10^{-11}$.\,
As these measurements, notably the $K^+$ ones, have moved increasingly close to their SM
predictions, the room for NP in \,$K\to\pi\slashed E$\, has become quite small.

As it turns out, of the possible underlying \,$s\to d\slashed E$\, operators
\cite{Su:2019tjn,Badin:2010uh,Kamenik:2011vy}, these decays are sensitive to only a subset.
Specifically, they can probe four-particle operators that have parity-even $ds$ quark bilinears
but are unaffected by those with exclusively parity-odd $ds$
bilinears~\cite{Tandean:2019tkm,Su:2019tjn,Li:2019cbk,Kamenik:2011vy}.
However, the latter operators can contribute to kaon reactions emitting no or two pions, namely
\,$K\to\slashed E$\, and \,$K\to\pi\pi\slashed E$,\, as well as to analogous decays
in the hyperon sector~\cite{Tandean:2019tkm,Su:2019tjn,Li:2019cbk}.
This means that, since at the moment there are precious few data on these
processes~\cite{Zyla:2020zbs}, searches for them might still come up with substantial
manifestations of NP or at least yield useful information about it complementary to that
supplied by \,$K\to\pi\slashed E$\, measurements.

In this paper, we adopt a model-independent approach to explore how big the branching fractions
of the various \,$K\to\pi\pi\slashed E$\, modes might be, taking into account the available
pertinent constraints.
We assume especially that the invisibles comprise a pair of spin-1/2 fermions or spinless
bosons, all of which are singlets under the SM gauge groups.
It is hoped that the outcomes of our study will motivate renewed attempts to pursue these decays
as NP tests.

The organization of the rest of the article is the following.
In Sec.\,\ref{npint}, we describe the quark-level operators responsible for the interactions of
interest.
In Sec.\,\ref{amplitudes}, we derive the amplitudes for the aforementioned kaon decay modes and
calculate their rates.
For the majority of them, we also write down the corresponding numerical branching fractions
in terms of the coefficients of the operators.
In Sec.\,\ref{smB}, we compare the SM predictions for these transitions with their current data.
In Sec.\,\ref{bsm}, we address the allowed maximal branching fractions of
\,$K\to\pi\pi\slashed E$\, due to NP and present our conclusions.
In all the instances we will discuss, we take the invisibles to be light enough that their masses
can be neglected compared to those of the mesons, which helps maximize the kaon decay rates.

\section{Interactions\label{npint}}

Depending on the types of particles carrying away the missing energy, the effective
$s\to d\slashed E$ operators are generally subject to different sets of restrictions.
If the invisible particles are SM neutrinos, which have charged-lepton partners because of the SM
$SU(2)_L$-gauge invariance, the operators would likely have to face stringent restraints from
lepton-flavor violation data.
Since these do not apply if the invisibles are SM-gauge singlets, hereafter we consider a couple
of cases involving them.

The missing energy is carried away by a pair of spin-1/2 Dirac fermions, \texttt f and
$\texttt f'$, in the first scenario and by a pair of complex spin-0 bosons, $\phi$ and $\phi'$,
in the second one.\footnote{In the recent literature covering the impact of NP on
\,$K\to\pi\pi\slashed E$, there are other possibilities for what carries away the missing energy.
In particular, it could alternatively be due to a single particle such as a massless dark
photon~\cite{Fabbrichesi:2017vma,Su:2019ipw,Su:2020xwt} or an invisible
axion~\cite{MartinCamalich:2020dfe}.}
At low energies, the relevant quark-level operators need to respect the strong and electromagnetic
gauge symmetries and are mostly obtainable from
the literature~\cite{Su:2019tjn,Badin:2010uh,Kamenik:2011vy}.
We can express the effective interaction Lagrangians as
\begin{align} \label{Lf}
{\cal L}_{\texttt{ff}'}^{} & \,=\, -\Big[ \overline d\gamma^\eta s\;
\overline{\texttt f} \gamma_\eta^{} \big( {\tt C}_{\texttt{ff}'}^{\textsc v}
+ \gamma_5^{} {\tt C}_{\texttt{ff}'}^{\textsc a} \big) {\texttt f}'
+ \overline d s\; \overline{\texttt f} \big( {\tt C}_{\texttt{ff}'}^{\textsc s}
+ \gamma_5^{} {\tt C}_{\texttt{ff}'}^{\textsc p} \big) {\texttt f}'
+ \overline d \sigma^{\eta\kappa}s\; \overline{\texttt f} \sigma_{\eta\kappa}^{} \big(
{\tt C}_{\texttt{ff}'}^{\textsc t} + \gamma_5^{} {\tt C}_{\texttt{ff}'}^{\textsc t\prime} \big)
{\texttt f}'
\nonumber \\ & ~~~~ ~~~~ +\,
\overline d \gamma^\eta\gamma_5^{} s\; \overline{\texttt f} \gamma_\eta^{} \big(
\tilde{c}{}_{\texttt{ff}'}^{\textsc v} + \gamma_5^{}
\tilde{c}{}_{\texttt{ff}'}^{\textsc a} \big) {\texttt f}'
+ \overline d \gamma_5^{} s\; \overline{\texttt f} \big(
\tilde{c}{}_{\texttt{ff}'}^{\textsc s} + \gamma_5^{}
\tilde{c}{}_{\texttt{ff}'}^{\textsc p} \big) {\texttt f}' \Big]
+\, {\rm H.c.} &
\end{align}
and
\begin{align} \label{Lphi}
{\cal L}_{\phi\phi'}^{} \,=\, - \Big( c_{\phi\phi'\,}^{\textsc v} \overline d\gamma^\eta s
+ c_{\phi\phi'\,}^{\textsc a} \overline d\gamma^\eta\gamma_5^{}s \Big)
i\Big( \phi^\dagger\partial_\eta^{}\phi' - \partial_\eta^{}\phi^\dagger\phi' \Big)
- \Big( c_{\phi\phi'\,}^{\textsc s} \overline ds + c_{\phi\phi'\,}^{\textsc p}
\overline d\gamma_5^{}s \Big) \phi^\dagger \phi' \,+\, {\rm H.c.}
\end{align}
for the two scenarios, respectively, where \,$\sigma^{\eta\kappa}=i[\gamma^\eta,\gamma^\kappa]/2$\,
and the $\texttt C$s, $\tilde{c}$s, and $c$s are in general complex coefficients which have
the dimension of inverse squared mass, except for $c_{\phi\phi'}^{\textsc s,\textsc p}$
which are of inverse-mass dimension.
These are free parameters in our model-independent approach and will be treated phenomenologically
in our numerical work later on.
In Eq.\,(\ref{Lf}) there are merely two tensor operators due to the identity
\,$2i\sigma^{\alpha\omega}\gamma_5^{}=\epsilon^{\alpha\omega\beta\psi}\sigma_{\beta\psi}$.
If \,$\texttt f'\neq\texttt f$ ($\phi'\neq\phi$), we implicitly also have another Lagrangian,
${\cal L}_{\texttt f'\texttt f}$ (${\cal L}_{\phi'\phi}$), which is the same as
${\cal L}_{\texttt{ff}'}$ (${\cal L}_{\phi\phi'}$) but with {\texttt f} and $\texttt f'$
($\phi$ and $\phi'$) interchanged.
We note that ${\cal L}_{\texttt{ff}'}$ and ${\cal L}_{\phi\phi'}$ could originate from
Lagrangians that are invariant under all the SM gauge groups~\cite{Su:2019tjn,Kamenik:2011vy}.

\section{Decay amplitudes and rates\label{amplitudes}}

To examine the amplitudes for the kaon decays of concern, we need the mesonic matrix elements
of the quark portions of the operators in Eqs.\,\,(\ref{Lf}) and (\ref{Lphi}).
They can be estimated with the aid of flavor-SU(3) chiral perturbation theory at
leading order \cite{Kamenik:2011vy,Tandean:2019tkm,He:2005we}.
For \,$K_{L,S}\to\slashed E$, the relevant hadronic matrix elements are
\begin{align} \label{<0K>}
\langle0|\overline{d}\gamma^\alpha\gamma_5^{}s|\,\overline{\!K}{}^0\rangle & \,=\,
\langle0|\overline{s}\gamma^\alpha\gamma_5^{}d|K^0\rangle \,=\, -if_K^{}p_K^\alpha \,, ~~~ ~~
\nonumber \\
\langle0|\overline{d}\gamma_5^{}s|\,\overline{\!K}{}^0\rangle & \,=\,
\langle0|\overline{s}\gamma_5^{}d|K^0\rangle \,=\, i B_0^{}f_K^{} \,,
\end{align}
with \,$f_K^{}\simeq156$\,MeV \cite{Zyla:2020zbs} being the kaon decay constant and
\,$B_0=m_K^2/(\bar m+m_s)\simeq2.0$\,\,GeV\, involving the average kaon mass and the combination
\,$\bar m+m_s\simeq124$\,\,MeV\, of light-quark masses at a renormalization scale
of 1 GeV, while for \,$K\to\pi\slashed E$,
\begin{align} \label{<piK>}
\langle\pi^-|\overline d\gamma^\alpha s|K^-\rangle & \,=\, p_K^\alpha+p_\pi^\alpha \,, ~~~~~~~
\langle\pi^-|\overline d s|K^-\rangle \,=\, B_0^{} \,, ~~~
\nonumber \\
\langle\pi^-|\overline d\sigma^{\alpha\kappa}s|K^-\rangle & \,=\,
2i a_T^{} \big( p_\pi^\alpha p_K^\kappa-p_\pi^\kappa p_K^\alpha \big) \,,
\end{align}
where $p_K^{}$ and $p_\pi^{}$ denote the kaon and pion momenta, respectively,
and $a_T$ is a constant having the dimension of inverse mass.
Assuming isospin symmetry and making use of charge conjugation, we further have
\,$\langle\pi^0|\overline d(\gamma^\eta,1,\sigma^{\eta\kappa})s|\,\overline{\!K}{}^0\rangle =
\langle\pi^0|\overline s(-\gamma^\eta,1,-\sigma^{\eta\kappa})d|K^0\rangle =
-\langle\pi^-|\overline d(\gamma^\eta,1,\sigma^{\eta\kappa})s|K^-\rangle/\sqrt2$.\,
For $K\to\pi\pi\slashed E$, we find~\cite{Tandean:2019tkm,Kamenik:2011vy,Su:2020xwt}
\begin{align} \label{<K->pp>} & \begin{array}[b]{rl}
\big\langle\pi^0(p_0^{})\,\pi^-(p_-^{})\big|\overline d(\gamma^\eta,1)\gamma_5^{}s
\big|K^-\big\rangle
\,= & \displaystyle \frac{i\sqrt2}{f_K^{}} \bigg[ \big( p_0^\eta - p_-^\eta, 0 \big)
+ \frac{\big(p_0^\alpha-p_-^\alpha\big) \tilde q_\alpha^{}}{m_K^2-\tilde q{}^2}
\big( \tilde q{}^\eta, -B_0^{} \big) \bigg] , ~~~
\bigskip \\
\big\langle\pi^+(p_+)\,\pi^-(p_-^{})\big|\overline d(\gamma^\eta,1)\gamma_5^{}s
\big|\,\overline{\!K}{}^0\big\rangle
& \displaystyle =\, \frac{2i}{f_K^{}} \bigg[ \big( p_+^\eta , 0 \big)
+ \frac{p_+^\alpha\, \tilde q_\alpha^{}}{m_K^2-\tilde q{}^2}
\big( \tilde q{}^\eta, -B_0^{} \big) \bigg] ,
\medskip \\
\big\langle\pi^+(p_+)\,\pi^-(p_-^{})\big|\overline s(\gamma^\eta,1)\gamma_5^{}d\big|K^0\big\rangle
&\displaystyle =\, \frac{2i}{f_K^{}} \bigg[ \big( p_-^\eta , 0 \big)
+ \frac{p_-^\alpha\, \tilde q_\alpha^{}}{m_K^2-\tilde q{}^2}
\big( \tilde q{}^\eta, -B_0^{} \big) \bigg] ,
\bigskip \\
\big\langle\pi^0(p_1^{})\,\pi^0(p_2^{})\big|\overline d(\gamma^\eta,1)\gamma_5^{}s
\big|\,\overline{\!K}{}^0\big\rangle
\,= & \big\langle\pi^0(p_1^{})\,\pi^0(p_2^{})\big|\overline s(\gamma^\eta,1)\gamma_5^{}d
\big|K^0\big\rangle
\vspace{3pt} \\ \,= & \displaystyle
\frac{i}{f_K^{}} \bigg[ \big( p_1^\eta + p_2^\eta, 0 \big)
+ \frac{\big(p_1^\alpha+p_2^\alpha\big) \tilde q_\alpha^{}}{m_K^2-\tilde q{}^2}
\big( \tilde q{}^\eta, -B_0^{} \big) \bigg] , \end{array}
\end{align}
\vspace{-3ex}
\begin{align} \label{<Kbz2pipi'>}
\langle\pi^0(p_0^{})\, \pi^-(p_-^{})|\overline d\sigma_{\eta\kappa}^{}s|K^-\rangle & \,=\,
\frac{i\sqrt2\, a_T^{}}{f_K^{}}\, \epsilon_{\eta\kappa\mu\tau}^{}
\big[ 4p_-^\mu p_0^\tau + \big(p_-^\mu-p_0^\mu\big)\tilde q^\tau \big] \,,
\nonumber \\
\big\langle\pi^+(p_+)\,\pi^-(p_-^{})\big|\overline d\sigma_{\eta\kappa}^{}s
\big|\,\overline{\!K}{}^0\big\rangle
& \,=\, \frac{2i a_T^{}}{f_K^{}}\, \epsilon_{\eta\kappa\mu\tau}^{}
\big(2p_-^\mu+\tilde q^\mu\big) p_+^\tau \,,
\nonumber \\
\big\langle\pi^+(p_+)\,\pi^-(p_-^{})\big|\overline s\sigma_{\eta\kappa}^{}d\big|K^0\big\rangle
& \,=\, \frac{2i a_T^{}}{f_K^{}}\, \epsilon_{\eta\kappa\mu\tau}^{}\,
p_-^\mu \big(2p_+^\tau+\tilde q^\tau\big) \,,
\nonumber \\
\big\langle\pi^0(p_1^{})\,\pi^0(p_2^{})\big|\overline d\sigma_{\eta\kappa}^{}s
\big|\,\overline{\!K}{}^0\big\rangle
& \,=\, -\big\langle\pi^0(p_1^{})\,\pi^0(p_2^{})\big|\overline s\sigma_{\eta\kappa}^{}d
\big|K^0\big\rangle
\,=\, \frac{ia_T^{}}{f_K^{}}\, \epsilon_{\eta\kappa\mu\tau}^{}\, \tilde q^\mu
\big(p_1^\tau+p_2^\tau\big) \,, ~~~
\end{align}
where
\,$\tilde q=p_K^{}-p_0^{}-p_-^{}=p_K^{}-p_-^{}-p_+^{}=p_K^{}-p_1^{}-p_2^{}$.\,
Although generally the matrix elements in Eqs.\,(\ref{<piK>})-(\ref{<Kbz2pipi'>}) involve
momentum-dependent form factors, to investigate the NP influence on these processes in this
study we do not need a high degree of precision and therefore can disregard form-factor effects.
We also ignore $\langle\pi^{0,-}\pi^+|\overline s\gamma^\eta d|K^{+,0}\rangle$, and their charge
conjugates, as they arise from small contributions derived from the anomaly Lagrangian, which
occurs at next-to-leading order in the chiral expansion~\cite{Geng:1994cw,Kamenik:2011vy}.

We now apply these matrix elements to kaon decays induced by ${\cal L}_{\texttt{ff}'}$ in
Eq.\,(\ref{Lf}) and take the {\texttt f} and $\texttt f'$ masses to be negligible,
$i.e.$ \,$m_{\texttt f,\texttt f'}\simeq0$.\,
Thus, for \,$K_{L,S}\to{\texttt f}\bar{\texttt f}{}'$,\, with the approximate relations
\,$\sqrt2\, K_{L,S}=K^0\pm\,\overline{\!K}{}^0$,\, we obtain the amplitudes to be
\begin{align} \label{MK2ff}
{\cal M}_{K_{L,S\,}^{}\to{\texttt f}\bar{\texttt f}{}'}^{} & \,=\, \tfrac{i}{\sqrt2} B_0^{}
f_K^{}\, \bar u_{\texttt f}^{} \Big( \tilde{\tt S}_{K_{L,S}^{}\texttt{ff}'}^{}
+ \gamma_5^{}\tilde{\tt P}_{K_{L,S}^{}\texttt{ff}'}^{} \Big)
v_{\bar{\texttt f}{}'}^{} \,, &
\end{align}
from which follow the decay rates
\begin{align} \label{GK2ff}
\Gamma_{K_{L,S\,}\to{\texttt f}\bar{\texttt f}{}'}^{} & \,=\,
\frac{B_0^2 f_K^2 m_{K^0}^{}}{16\pi} \Big(
\big|\tilde{\tt S}_{K_{L,S}^{}\texttt{ff}'} \big| \raisebox{2pt}{$^2$}
+ \big|\tilde{\tt P}_{K_{L,S}^{}\texttt{ff}'} \big| \raisebox{2pt}{$^2$} \Big) \,, &
\end{align}
where
\begin{align} \label{SK2ff}
\tilde{\tt S}_{K\!_L^{}\texttt{ff}'}^{} & = \tilde{c}_{\texttt{ff}'}^{\textsc s}
- \tilde{c}_{\texttt f'\texttt f}^{\textsc s*} \,, &
\tilde{\tt P}_{K\!_L^{}\texttt{ff}'}^{} & = \tilde{c}_{\texttt{ff}'}^{\textsc p}
+ \tilde{c}_{\texttt f'\texttt f}^{\textsc p*} \,, \nonumber \\
\tilde{\tt S}_{K\!_S^{}\texttt{ff}'}^{} & = -\tilde{c}_{\texttt{ff}'}^{\textsc s}
- \tilde{c}_{\texttt f'\texttt f}^{\textsc s*} \,, &
\tilde{\tt P}_{K\!_S^{}\texttt{ff}'}^{} & = \tilde{c}_{\texttt f'\texttt f}^{\textsc p*}
- \tilde{c}_{\texttt{ff}'}^{\textsc p} \,. &
\end{align}
We can see that \,$K_{L,S}\to{\texttt f}\bar{\texttt f}{}'$\, are insensitive to
$\texttt C^{\textsc v,\textsc a,\textsc s,\textsc p,\textsc t,\textsc t\prime}$ and
$\tilde{c}^{\textsc v}$ as well as $\tilde{c}^{\textsc a}$ if \,$m_{\texttt f,\texttt f'}=0$.\,

For \,$K\to\pi{\texttt f}\bar{\texttt f}{}'$,\, we express the amplitude as
\,${\cal M}_{K\to\pi{\texttt f}\bar{\texttt f}{}'} =
\bar u_{\texttt f}^{} \big( S_{K\pi\texttt{ff}'}^{}
+ P_{K\pi\texttt{ff}'\,}^{} \gamma_5^{} \big) v_{\bar{\texttt f}{}'}^{}$.\,
The $S$ and $P$ terms for \,$K^-\to\pi^-{\texttt f}\bar{\texttt f}{}'$\, and
\,$K_L\to\pi^0{\texttt f}\bar{\texttt f}{}'$\, are
\begin{align} \label{SPK2pff}
S_{K^-\pi^-\texttt{ff}'}^{} & \,=\, 2\, \slashed p{}_K^{}\, \texttt C_{\texttt{ff}'}^{\textsc v}
+ B_0^{}\, \texttt C_{\texttt{ff}'}^{\textsc s}
+ 4 a_T^{}\, p_K^{}\!\cdot\!\big(p_{\bar{\texttt f}{}'}^{}-p_{\texttt f}^{}\big)\,
{\tt C}_{\texttt{ff}'}^{\textsc t} \,,
\nonumber \\
P_{K^-\pi^-\texttt{ff}'}^{} & \,=\, 2\, \slashed p{}_K^{}\, \texttt C_{\texttt{ff}'}^{\textsc a}
+ B_0^{}\, \texttt C_{\texttt{ff}'}^{\textsc p}
+ 4 a_T^{}\, p_K^{}\!\cdot\!\big(p_{\bar{\texttt f}{}'}^{}-p_{\texttt f}^{}\big)\,
{\tt C}_{\texttt{ff}'}^{\textsc t\prime} \,,
\nonumber \\
S_{K_L^{}\pi^0\texttt{ff}'}^{} & \,=\, \big( \texttt C_{\texttt f'\texttt f}^{\textsc v*}
- \texttt C_{\texttt{ff}'}^{\textsc v} \big) \slashed p{}_K^{}
- \tfrac{1}{2} B_0^{} \big( \texttt C_{\texttt f'\texttt f}^{\textsc s*}
+ \texttt C_{\texttt{ff}'}^{\textsc s} \big)
+ 2 a_T^{}\,p_K^{}\!\cdot\!\big(p_{\bar{\texttt f}{}'}^{}-p_{\texttt f}^{}\big) \big(
{\tt C}_{\texttt f'\texttt f}^{\textsc t*} - {\tt C}_{\texttt{ff}'}^{\textsc t} \big) \,, ~~~
\nonumber \\
P_{K_L^{}\pi^0\texttt{ff}'}^{} & \,=\, \big( \texttt C_{\texttt f'\texttt f}^{\textsc a*}
- \texttt C_{\texttt{ff}'}^{\textsc a} \big) \slashed p{}_K^{}
+ \tfrac{1}{2} B_0^{} \big( \texttt C_{\texttt f'\texttt f}^{\textsc p*}
- \texttt C_{\texttt{ff}'}^{\textsc p} \big)
- 2 a_T^{}\,p_K^{}\!\cdot\!\big(p_{\bar{\texttt f}{}'}^{}-p_{\texttt f}^{}\big) \big(
{\tt C}_{\texttt f'\texttt f}^{\textsc t\prime*} + {\tt C}_{\texttt{ff}'}^{\textsc t\prime}
\big) \,.
\end{align}
These lead to the differential rates
\begin{align} \label{G'KL2pff} & \begin{array}[t]{l} \displaystyle
\frac{d\Gamma_{K^-\to\pi^-{\texttt f}\bar{\texttt f}{}'}}{d\hat s} =
\frac{\lambda_{K^-\pi^-}^{3/2}}{192\pi^3 m_{K^-}^3} \Bigg[
|\texttt C_{\texttt{ff}'}^{\textsc v}|^2 + |\texttt C_{\texttt{ff}'}^{\textsc a}|^2
+ 3 B_0^2 \hat s\, \frac{|\texttt C_{\texttt{ff}'}^{\textsc s}|^2
+ |\texttt C_{\texttt{ff}'}^{\textsc p}|^2}{2\lambda_{K^-\pi^-}}
+ 2 a_T^2 \big( |\texttt C_{\texttt{ff}'}^{\textsc t}|^2
+ |\texttt C_{\texttt{ff}'}^{\textsc t\prime}|^2 \big) \hat s \Bigg] \,, \end{array}
\nonumber \\ & \begin{array}[b]{rl} \displaystyle
\frac{d\Gamma_{K_L^{}\to\pi^0{\texttt f}\bar{\texttt f}{}'}}{d\hat s} =
\frac{\lambda_{K^0\pi^0}^{3/2}}{768\pi^3 m_{K^0}^3} & \displaystyle \!\!\!
\Bigg[ |\texttt C_{\texttt{ff}'}^{\textsc v} - \texttt C_{\texttt f'\texttt f}^{\textsc v*}|^2
+ |\texttt C_{\texttt{ff}'}^{\textsc a}-\texttt C_{\texttt f'\texttt f}^{\textsc a*}|^2
+ 3 B_0^2 \hat s\, \frac{|\texttt C_{\texttt{ff}'}^{\textsc s}
+ \texttt C_{\texttt f'\texttt f}^{\textsc s*}|^2 + |\texttt C_{\texttt{ff}'}^{\textsc p}
- \texttt C_{\texttt f'\texttt f}^{\textsc p*}|^2}{2\lambda_{K^0\pi^0}}
\\ & \displaystyle +\;
2 a_T^2 \big( |\texttt C_{\texttt{ff}'}^{\textsc t}
- \texttt C_{\texttt f'\texttt f}^{\textsc t*}|^2 + |\texttt C_{\texttt{ff}'}^{\textsc t\prime}
+ \texttt C_{\texttt f'\texttt f}^{\textsc t\prime*}|^2 \big) \hat s \Bigg] \,, \end{array}
\end{align}
where $\hat s$ represents the invariant mass squared of the ${\texttt f}\bar{\texttt f}{}'$ pair,
\begin{align}
\lambda_{AB}^{} & \,=\,
{\cal K}\big(m_A^2,m_B^2,\hat s\big) \,, &
{\cal K}(x,y,z) & \,=\, (x-y-z)^2-4yz \,. ~~~
\end{align}
Evidently, \,$K\to\pi{\texttt f}\bar{\texttt f}{}'$,\, unlike
\,$K\to{\texttt f}\bar{\texttt f}{}'$,\, can probe
$\texttt C^{\textsc v,\textsc a,\textsc s,\textsc p,\textsc t,\textsc t\prime}$,
but not $\tilde{c}^{\textsc v,\textsc a,\textsc s,\textsc p}$.

For \,$K^-\to\pi^0\pi^-{\texttt f}\bar{\texttt f}{}'$\, and
\,$K_L\to(\pi^+\pi^-,\pi^0\pi^0){\texttt f}\bar{\texttt f}{}'$,\, we find
\begin{align} \nonumber
{\cal M}_{K^-\to\pi^0\pi^-{\texttt f}\bar{\texttt f}{}'}^{} \,=\, \frac{i\sqrt2}{f_K^{}}\,
\bar u_{\texttt f}^{} \! & \begin{array}[t]{l} \displaystyle \bigg\{
\big(\slashed p{}_0^{}-\slashed p{}_-^{}\big) \big( \tilde{c}_{\texttt{ff}'}^{\textsc v}
+ \gamma_5^{} \tilde{c}_{\texttt{ff}'}^{\textsc a} \big)
+ \frac{B_0}{\tilde{\textsc k}{}^2} \big( \tilde{c}_{\texttt{ff}'}^{\textsc s}
+ \gamma_5^{} \tilde{c}_{\texttt{ff}'}^{\textsc p} \big)
\big(p_-^\alpha-p_0^\alpha\big) \hat q_\alpha^{}
\\ \,+~
2i a_T^{} \big[ 4 p_-^\alpha p_0^\tau + \big(p_-^\alpha-p_0^\alpha\big) \hat q^\tau \big]
\sigma_{\alpha\tau}^{} \big( \gamma_5^{} {\tt C}_{\texttt{ff}'}^{\textsc t}
+ {\tt C}_{\texttt{ff}'}^{\textsc t\prime} \big)
\bigg\} v_{\bar{\texttt f}{}'}^{} \,, \end{array}
\nonumber \\
{\cal M}_{K_L\to\pi^+\pi^-{\texttt f}\bar{\texttt f}{}'}^{} \,=\,
\frac{i\sqrt2}{f_K^{}}\, \bar u_{\texttt f}^{} \! & \begin{array}[t]{l} \displaystyle \bigg[
\slashed p_+ \big( \tilde{c}_{\texttt{ff}'}^{\textsc v} + \gamma_5^{}
\tilde{c}_{\texttt{ff}'}^{\textsc a} \big) - \frac{B_0}{\tilde{\textsc k}{}^2}
\big( \tilde{c}_{\texttt{ff}'}^{\textsc s} + \gamma_5^{}
\tilde{c}_{\texttt{ff}'}^{\textsc p} \big)\, p_+^\alpha\, \hat q_\alpha^{}
\smallskip \\ \displaystyle \,+~
2i a_T^{}\, \big(2p_-^\alpha+\hat q^\alpha\big) p_+^\tau\, \sigma_{\alpha\tau}^{}
\big(\gamma_5^{} {\tt C}_{\texttt{ff}'}^{\textsc t} + {\tt C}_{\texttt{ff}'}^{\textsc t\prime}\big)
\smallskip \\ \displaystyle \,+~
\slashed p_- \big( \tilde{c}_{\texttt f'\texttt f}^{\textsc v*} + \gamma_5^{}
\tilde{c}_{\texttt f'\texttt f}^{\textsc a*} \big) + \frac{B_0}{\tilde{\textsc k}{}^2}
\big( \tilde{c}_{\texttt f'\texttt f}^{\textsc s*} - \gamma_5^{}
\tilde{c}_{\texttt f'\texttt f}^{\textsc p*} \big)\, p_-^\alpha\, \hat q_\alpha^{}
\\ \,+~
2i a_T^{}\, p_-^\alpha \big(2p_+^\tau+\hat q^\tau\big)\, \sigma_{\alpha\tau}^{} \big( \gamma_5^{}
{\tt C}_{\texttt f'\texttt f}^{\textsc t*} - {\tt C}_{\texttt f'\texttt f}^{\textsc t\prime*}
\big) \bigg] v_{\bar{\texttt f}{}'}^{} \,, \end{array} &
\nonumber \\
{\cal M}_{K_L\to\pi^0\pi^0{\texttt f}\bar{\texttt f}{}'}^{} \,=\, \frac{i}{\sqrt2 f_K^{}}\,
\bar u_{\texttt f}^{} \! & \begin{array}[t]{l} \displaystyle \bigg\{ \slashed p_K \big[
\tilde{c}_{\texttt{ff}'}^{\textsc v} + \tilde{c}_{\texttt f'\texttt f}^{\textsc v*}
+ \gamma_5^{} \big( \tilde{c}_{\texttt{ff}'}^{\textsc a}
+ \tilde{c}_{\texttt f'\texttt f}^{\textsc a*} \big) \big]
\\ \displaystyle \,+~
\frac{B_0}{\tilde{\textsc k}{}^2} \big[ \tilde{c}_{\texttt{ff}'}^{\textsc s}
- \tilde{c}_{\texttt f'\texttt f}^{\textsc s*} + \gamma_5^{} \big(
\tilde{c}_{\texttt{ff}'}^{\textsc p}
+ \tilde{c}_{\texttt f'\texttt f}^{\textsc p*} \big) \big]
\big(\hat s-p_K^\alpha \hat q_\alpha^{}\big)
\\ \displaystyle \,+~
2i a_T^{}\, \hat q^\alpha p_K^\tau\, \sigma_{\alpha\tau}^{} \big[ \gamma_5^{}
\big({\tt C}_{\texttt{ff}'}^{\textsc t}-{\tt C}_{\texttt f'\texttt f}^{\textsc t*}\big)
+ {\tt C}_{\texttt{ff}'}^{\textsc t\prime} + {\tt C}_{\texttt f'\texttt f}^{\textsc t\prime*}
\big] \bigg\} v_{\bar{\texttt f}}^{} \,, \end{array} &
\end{align}
where
\begin{align}
\hat q & \,=\, p_{\texttt f}^{}+p_{\bar{\texttt f}{}'}^{} \,, &
\hat s & \,=\, \hat q^2 \,, &
\tilde{\textsc k}{}^2 & \,=\, m_K^2-\hat s \,, &
\end{align}
with $m_K$ in $\tilde{\textsc k}$ being the average kaon mass.
We then arrive at the double differential rates
\begin{subequations} \label{G''K2pp'ff}
\begin{align}
\frac{d^2\Gamma_{K^-\to\pi^0\pi^-{\texttt f}\bar{\texttt f}{}'}}{d\hat s\,d\hat\varsigma} =
\frac{\beta_{\hat\varsigma}^3 \tilde\lambda{}_{K^-}^{3/2}}{(4\pi)^5 f_K^2} & \Bigg\{
\Bigg(1+\frac{12\hat s\hat\varsigma}{\tilde\lambda_{K^-}}\Bigg)
\frac{|\tilde{c}_{\texttt{ff}'}^{\textsc v}|^2 +
|\tilde{c}_{\texttt{ff}'}^{\textsc a}|^2}{18\, m_{K^-}^3}
+ B_0^2\hat s\, \frac{|\tilde{c}_{\texttt{ff}'}^{\textsc s}|^2
+ |\tilde{c}_{\texttt{ff}'}^{\textsc p}|^2}{12\, \tilde{\textsc k}{}^4\, m_{K^-}^3}
\nonumber \\ & \,+\,
a_T^2 \Bigg[ \hat s+4\hat\varsigma + \frac{12 \hat\varsigma}{\tilde\lambda_{K^-}}
\big(m_{K^-}^2-\hat\varsigma\big)^{\!2} \Bigg] \frac{ |\texttt C_{\texttt{ff}'}^{\textsc t}|^2
+ |\texttt C_{\texttt{ff}'}^{\textsc t\prime}|^2}{9\,m_{K^-}^3} \Bigg\} \,,
\end{align}
\begin{align}
\frac{d^2\Gamma_{K_L\to\pi^+\pi^-{\texttt f}\bar{\texttt f}{}'}}{d\hat s\,d\hat\varsigma} =
\frac{\beta_{\hat\varsigma}^3 \tilde\lambda{}_{K^0}^{3/2}}{4(4\pi)^5 f_K^2} & \Bigg\{
\Bigg(1+\frac{12\hat s\hat\varsigma}{\tilde\lambda_{K^0}}\Bigg)
\frac{|\tilde{c}_{\texttt{ff}'}^{\textsc v}
- \tilde{c}_{\texttt f'\texttt f}^{\textsc v*}|^2
+ |\tilde{c}_{\texttt{ff}'}^{\textsc a}
- \tilde{c}_{\texttt f'\texttt f}^{\textsc a*}|^2}{18\, m_{K^0}^3}
\nonumber \\ & \,+\,
B_0^2 \hat s\, \frac{|\tilde{c}_{\texttt{ff}'}^{\textsc s}
+ \tilde{c}_{\texttt f'\texttt f}^{\textsc s*}|^2
+ |\tilde{c}_{\texttt{ff}'}^{\textsc p}
- \tilde{c}_{\texttt f'\texttt f}^{\textsc p*}|^2}{12\, \tilde{\textsc k}{}^4\, m_{K^0}^3}
+ \frac{|\tilde{c}_{\texttt{ff}'}^{\textsc v}
+ \tilde{c}_{\texttt f'\texttt f}^{\textsc v*}|^2
+ |\tilde{c}_{\texttt{ff}'}^{\textsc a}
+ \tilde{c}_{\texttt f'\texttt f}^{\textsc a*}|^2}{6 \beta_{\hat\varsigma}^2\, m_{K^0}^3}
\nonumber \\ & \,+\,
\Bigg(1+\frac{4\hat s\hat\varsigma}{\tilde\lambda_{K^0}}\Bigg) B_0^2 \hat s\,
\frac{|\tilde{c}_{\texttt{ff}'}^{\textsc s}
- \tilde{c}_{\texttt f'\texttt f}^{\textsc s*}|^2
+ |\tilde{c}_{\texttt{ff}'}^{\textsc p}
+ \tilde{c}_{\texttt f'\texttt f}^{\textsc p*}|^2}
{4 \beta_{\hat\varsigma}^2\, \tilde{\textsc k}{}^4\, m_{K^0}^3}
\nonumber \\ & \,+\,
a_T^2 \Bigg[ \hat s+4\hat\varsigma + \frac{12 \hat\varsigma}{\tilde\lambda_{K^0}}
\big(m_{K^0}^2-\hat\varsigma\big)^{\!2} \Bigg] \frac{
|{\tt C}_{\texttt{ff}'}^{\textsc t}+{\tt C}_{\texttt f'\texttt f}^{\textsc t*}|^2
+ |{\tt C}_{\texttt{ff}'}^{\textsc t\prime}-{\tt C}_{\texttt f'\texttt f}^{\textsc t\prime*}|^2}
{9\,m_{K^0}^3}
\nonumber \\ & \,+\, a_T^2 \hat s\,
\frac{|{\tt C}_{\texttt{ff}'}^{\textsc t}-{\tt C}_{\texttt f'\texttt f}^{\textsc t\prime*}|^2
+ |{\tt C}_{\texttt{ff}'}^{\textsc t}+{\tt C}_{\texttt f'\texttt f}^{\textsc t\prime*}|^2}
{3 \beta_{\hat\varsigma}^2\, m_{K^0}^3} \Bigg\} \,,
\end{align}
\begin{align}
\frac{d^2\Gamma_{K_L\to\pi^0\pi^0{\texttt f}\bar{\texttt f}{}'}}{d\hat s\,d\hat\varsigma} =
\frac{\beta_{\hat\varsigma}^{} \tilde\lambda_{K^0}^{3/2}}{8(4\pi)^5 f_K^2} & \Bigg[
\frac{|\tilde{c}_{\texttt{ff}'}^{\textsc v}
+ \tilde{c}_{\texttt f'\texttt f}^{\textsc v*}|^2
+ |\tilde{c}_{\texttt{ff}'}^{\textsc a}
+ \tilde{c}_{\texttt f'\texttt f}^{\textsc a*}|^2}{6\, m_{K^0}^3}
\nonumber \\ & \,+\,
B_0^2 \hat s \Bigg(1+\frac{4\hat s\hat\varsigma}{\tilde\lambda_{K^0}}\Bigg)
\frac{|\tilde{c}_{\texttt{ff}'}^{\textsc s}
- \tilde{c}_{\texttt f'\texttt f}^{\textsc s*}|^2
+ |\tilde{c}_{\texttt{ff}'}^{\textsc p}
+ \tilde{c}_{\texttt f'\texttt f}^{\textsc p*}|^2}
{4\, \tilde{\textsc k}{}^4\, m_{K^0}^3}
\nonumber \\ & \,+\,
a_T^2\hat s\,\frac{|{\tt C}_{\texttt{ff}'}^{\textsc t}-{\tt C}_{\texttt f'\texttt f}^{\textsc t*}|^2
+ |{\tt C}_{\texttt{ff}'}^{\textsc t\prime}+{\tt C}_{\texttt f'\texttt f}^{\textsc t\prime*}|^2}
{3\, m_{K^0}^3} \Bigg] \,, &
\end{align}
\end{subequations}
where $\hat\varsigma$ is the invariant mass squared of the pion pair,
\begin{align}
\beta_{\hat\varsigma}^{} & \,=\, \sqrt{1-\frac{4m_\pi^2}{\hat\varsigma}} \,, & &
\tilde\lambda_{\cal P}^{} \,=\,
{\cal K}\big(m_{\cal P}^2,\hat s,\hat\varsigma\big) \,, ~~~ ~~
{\cal P} \,=\, K^-, K^0 \,. ~~~
\end{align}
The $\hat s$ and $\hat\varsigma$ integration ranges for calculating the $K^-$ and $K_L$ partial
rates from Eq.\,(\ref{G''K2pp'ff}) are \,$0\le\hat s\le(m_{K^-,K^0}-2m_\pi)^2$\, and
\,$4m_\pi^2\le\hat\varsigma\le\big(m_{K^-,K^0}^{}-\hat s{}^{1/2}\big)\raisebox{1pt}{$^2$}$,\,
respectively.
For the mode with the $\pi^0\pi^-$ ($\pi^+\pi^-$ or $\pi^0\pi^0$) pair, $m_\pi$ refers to
the isospin-average (charged or neutral) pion mass.
The expressions for
the \,$K_S\to\pi\pi{\texttt f}\bar{\texttt f}{}'$\,
rates equal their $K_L$ counterparts computed from Eq.\,(\ref{G''K2pp'ff}) but with
the signs of $\tilde{c}_{\texttt{ff}'}^{\textsc v,\textsc a,\textsc s,\textsc p}$
and $\texttt C_{\texttt{ff}'}^{\textsc t,\textsc t\prime}$ flipped.
Clearly, \,$K\to\pi\pi{\texttt f}\bar{\texttt f}{}'$,\, as opposed to
\,$K\to\pi{\texttt f}\bar{\texttt f}{}'$,\, are sensitive to
$\tilde{c}^{\textsc v,\textsc a,\textsc s,\textsc p}$, besides
$\texttt C^{\textsc t,\textsc t\prime}$, but not to
$\texttt C^{\textsc v,\textsc a,\textsc s,\textsc p}$
in our approximation of the hadronic matrix elements.
We remark that the \,$\texttt f'=\texttt f$\, possibility has previously been considered in
Refs.\,\cite{Tandean:2019tkm,Kamenik:2011vy} and our formulas above applied to that case agree
with those given therein in the \,$m_{\texttt f}=0$\, limit.

If \,$\texttt f'\neq\texttt f$,\, the extra channels
\,$K\to(\pi^0,\pi^+\pi^-,\pi^0\pi^0)\texttt f'\bar{\texttt f}$\, also occur, whose rates are obtainable
from those of \,$K\to(\pi^0,\pi^+\pi^-,\pi^0\pi^0)\texttt f\bar{\texttt f}{}'$,\, respectively,
by interchanging the labels {\texttt f} and $\texttt f'$ of the coefficients.
For \,$\texttt f'\neq\texttt f$,\, if
$\tilde{c}_{\texttt f'\texttt f}^{\textsc v,\textsc a,\textsc s,\textsc p}$ and
$\texttt C_{\texttt f'\texttt f}^{\textsc t,\textsc t\prime}$ are not zero,
they are generally independent from
$\tilde{c}_{\texttt{ff}'}^{\textsc v,\textsc a,\textsc s,\textsc p}$ and
$\texttt C_{\texttt{ff}'}^{\textsc t,\textsc t\prime}$
and bring about \,$K^-\to\pi^0\pi^-\texttt f'\bar{\texttt f}$\, as well.

Turning to the processes induced by ${\cal L}_{\phi\phi'}$ in Eq.\,(\ref{Lphi}), we again assume
that the masses of the invisible particles, $\phi$ and $\phi'$, can be neglected,
\,$m_{\phi,\phi'}\simeq0$.\,
It follows that the amplitudes for \,$K_{L,S}\to\phi\bar\phi'$\, are
\begin{align} \label{MK->ss}
{\cal M}_{K_L\to\phi\bar\phi'}^{} & \,=\, \tfrac{i}{\sqrt2} \big( c_{\phi\phi'}^{\textsc p}
- c_{\phi'\phi}^{\textsc p*} \big) B_0^{} f_K^{} \,, &
{\cal M}_{K_S\to\phi\bar\phi'}^{} & \,=\, \tfrac{-i}{\sqrt2} \big( c_{\phi\phi'}^{\textsc p}
+ c_{\phi'\phi}^{\textsc p*} \big) B_0^{} f_K^{} \,,
\end{align}
which lead to the decay rates
\begin{align} \label{GK2ss}
\Gamma_{K_{L,S}\to\phi\bar\phi'}^{} & \,=\, \frac{B_0^2 f_K^2}{32\pi m_{K^0}^{}}
\big| c_{\phi\phi'}^{\textsc p} \mp c_{\phi'\phi}^{\textsc p*} \big|\raisebox{2pt}{$^2$} \,. ~~~
\end{align}
As for the two-body modes, we find
\begin{align} \label{G'K2pss}
\frac{d\Gamma_{K^-\to\pi^-\phi\bar\phi'}^{}}{d\hat s} & \,=\, \frac{\lambda_{K^-\pi^-}^{1/2}}
{768 \pi^3 m_{K^-}^3} \Big( \lambda_{K^-\pi^-}^{}
\big|c_{\phi\phi'}^{\textsc v}\big|\raisebox{1pt}{$^2$}
+ 3 B_0^2\, \big|c_{\phi\phi'}^{\textsc s}\big|\raisebox{1pt}{$^2$} \Big) \,,
\nonumber \\
\frac{d\Gamma_{K_L^{}\to\pi^0\phi\bar\phi'}^{}}{d\hat s} & \,=\, \frac{\lambda_{K^0\pi^0}^{1/2}}
{3072 \pi^3 m_{K^0}^3} \Big( \lambda_{K^0\pi^0}^{}
\big|c_{\phi\phi'}^{\textsc v}-c_{\phi'\phi}^{\textsc v*}\big|\raisebox{1pt}{$^2$} + 3 B_0^2\,
\big|c_{\phi\phi'}^{\textsc s}+c_{\phi'\phi}^{\textsc s*}\big|\raisebox{1pt}{$^2$} \Big) \,, &
\end{align}
where \,$\hat s=\hat{\texttt q}{}^2$\, with \,$\hat{\texttt q}=\texttt p+\bar{\texttt p}$\, being
the sum of the momenta \texttt p and $\bar{\texttt p}$ of $\phi$ and $\bar\phi'$, respectively.
For \,$K\to\pi\pi\phi\bar\phi'$,\, we derive
\begin{align}
{\cal M}_{K^-\to\pi^0\pi^-\phi\bar\phi'}^{} & =\, \frac{i\sqrt2\, \big(p_0^\tau-p_-^\tau\big)}
{f_K^{}} \Bigg[ c_{\phi\phi'}^{\textsc a} (\texttt p-\bar{\texttt p})_\tau^{}
- \frac{B_0^{}\, c_{\phi\phi'\,}^{\textsc p} \hat{\texttt q}{}_\tau^{}}{\tilde{\textsc k}{}^2}
\Bigg] \,,
\nonumber \\
{\cal M}_{K_L^{}\to\pi^+\pi^-\phi\bar\phi'}^{} & \,=\, \frac{i\sqrt2}{f_K^{}} \Bigg[ \big(
c_{\phi'\phi}^{\textsc a*}\, p_-^\tau + c_{\phi\phi'\,}^{\textsc a} p_+^\tau \big)
(\texttt p-\bar{\texttt p})_\tau^{} + \frac{B_0}{\tilde{\textsc k}{}^2} \big(
c_{\phi'\phi}^{\textsc p*}\, p_-^\tau - c_{\phi\phi'\,}^{\textsc p} p_+^\tau \big)
\hat{\texttt q}{}_\tau^{} \Bigg] \,,
\nonumber \\
{\cal M}_{K_L^{}\to\pi^0\pi^0\phi\bar\phi'}^{} & \,=\, \frac{i\big(p_1^\tau+p_2^\tau\big)}
{\sqrt2\, f_K^{}} \Bigg[ \big(c_{\phi'\phi}^{\textsc a*}+c_{\phi\phi'}^{\textsc a}\big)
(\texttt p-\bar{\texttt p})_\tau^{} + \frac{B_0}{\tilde{\textsc k}{}^2} \big(
c_{\phi'\phi}^{\textsc p*} - c_{\phi\phi'}^{\textsc p} \big) \hat{\texttt q}{}_\tau^{} \Bigg] \,, &
\end{align}
from which we arrive at
\begin{align} \label{G''K2pp'ss}
\frac{d^2\Gamma_{K^-\to\pi^0\pi^-\phi\bar\phi'}^{}}{d\hat s\,d\hat\varsigma} & \,=\,
\frac{4\beta_{\hat\varsigma}^3 \tilde{\lambda}_{K^-}^{1/2}}{3(8\pi)^5 f_K^2}
\Bigg[ \frac{\tilde\lambda_{K^-} + 12 \hat s \hat\varsigma}{3\, m_{K^-}^3}
\big|c_{\phi\phi'}^{\textsc a}\big|^2 + \frac{\tilde\lambda_{K^-} B_0^2}
{\tilde{\textsc k}{}^4\, m_{K^-}^3} \big|c_{\phi\phi'}^{\textsc p}\big|^2 \Bigg] \,,
\nonumber \\
\frac{d^2\Gamma_{K_L^{}\to\pi^+\pi^-\phi\bar\phi'}^{}}{d\hat s\,d\hat\varsigma} & \,=\,
\frac{\beta_{\hat\varsigma}^{~} \tilde{\lambda}_{K^0}^{1/2}}{3(8\pi)^5 f_K^2}
\!\! \begin{array}[t]{l} \displaystyle \Bigg[ \beta_{\hat\varsigma}^2
\frac{\tilde\lambda_{K^0} + 12 \hat s\hat\varsigma}{3\, m_{K^0}^3}
\big| c_{\phi'\phi}^{\textsc a*} - c_{\phi\phi'}^{\textsc a}\big|\raisebox{1pt}{$^2$}
+ \frac{\beta_{\hat\varsigma}^2 \tilde\lambda_{K^0} B_0^2}{\tilde{\textsc k}{}^4\, m_{K^0}^3}
\big|c_{\phi'\phi}^{\textsc p*} + c_{\phi\phi'}^{\textsc p}\big|\raisebox{1pt}{$^2$}
\\ \displaystyle \,+~
\frac{\tilde\lambda_{K^0}}{m_{K^0}^3} \big| c_{\phi'\phi}^{\textsc a*}
+ c_{\phi\phi'}^{\textsc a}\big|\raisebox{1pt}{$^2$} + \frac{3 \big( \tilde{\lambda}_{K^0}
+ 4\hat s\hat\varsigma\big) B_0^2}{\tilde{\textsc k}{}^4\, m_{K^0}^3}
\big|c_{\phi'\phi}^{\textsc p*} - c_{\phi\phi'}^{\textsc p}\big|\raisebox{1pt}{$^2$} \Bigg] \,,
\end{array} ~~~
\nonumber \\
\frac{d^2\Gamma_{K_L^{}\to\pi^0\pi^0\phi\bar\phi'}^{}}{d\hat s\,d\hat\varsigma} & \,=\,
\frac{\beta_{\hat\varsigma}^{~} \tilde{\lambda}_{K^0}^{1/2}}{6(8\pi)^5 f_K^2} \Bigg[
\frac{\tilde\lambda_{K^0}}{m_{K^0}^3} \big| c_{\phi'\phi}^{\textsc a*}
+ c_{\phi\phi'}^{\textsc a}\big|\raisebox{1pt}{$^2$}
+ \frac{3\big(\tilde\lambda_{K^0}+4\hat s\hat\varsigma\big)B_0^2}{\tilde{\textsc k}{}^4\,m_{K^0}^3}
\big|c_{\phi'\phi}^{\textsc p*} - c_{\phi\phi'}^{\textsc p}\big|\raisebox{1pt}{$^2$} \Bigg] \,.
\end{align}
The expressions for the \,$K_S\to\pi\pi\phi\bar\phi'$\, rates equal their $K_L$ counterparts
calculated from Eq.\,(\ref{G''K2pp'ss}) except that the signs of
$c_{\phi\phi'}^{\textsc a,\textsc p}$ are flipped.

The last paragraph shows that \,$K\to\phi\bar\phi'$\, are sensitive exclusively to $c^{\textsc p}$,
whereas \,$K\to\pi\pi\phi\bar\phi'$\, can probe solely the parity-odd couplings,
$c^{\textsc a}$ and $c^{\textsc p}$, in our approximation of the mesonic matrix elements.
By contrast, \,$K\to\pi\phi\bar\phi'$\, depend on the parity-even coefficients,
$c^{\textsc v}$~and $c^{\textsc s}$, but not on~$c^{\textsc a,\textsc p}$.
As in the fermionic scenario, if \,$\phi'\neq\phi$,\, the extra channels
\,$K\to(\pi^0,\pi^+\pi^-,\pi^0\pi^0)\phi'\bar\phi$\, also take place, as well as
\,$K^-\to\pi^0\pi^-\phi'\bar\phi$\, if \,$c_{\phi'\phi}\neq0$.\,
We comment that the \,$\phi'=\phi$\, possibility has also been considered before in
Refs.\,\cite{Li:2019cbk,Kamenik:2011vy} and our formulas above applied to that case are
consistent with those obtained therein for \,$m_\phi=0$.\,

\begin{table}[t]
\caption{Summary of coefficients in Eqs.\,\,(\ref{Lf}) and\,\,(\ref{Lphi}) contributing to
the various FCNC kaon decays with missing energy carried away by Dirac spin-1/2 fermions,
${\texttt f}\bar{\texttt f}{}'$, or by spin-0 bosons, $\phi\bar\phi'$, if their masses are
negligible, \,$m_{\texttt f,\texttt f',\phi,\phi'}\simeq0$.\label{couplings}}
\begin{tabular}{|c||c|c|c||c|c|c|} \hline
\small \,Decay modes\, & \,$K\to{\texttt f}\bar{\texttt f}{}'$\, &
\,$K\to\pi{\texttt f}\bar{\texttt f}{}'$\, &
\,$K\to\pi\pi{\texttt f}\bar{\texttt f}{}'\vphantom{\int_|^{|^|}}$\, &
\,$K\to\phi\bar\phi'$\, & \,$K\to\pi\phi\bar\phi'$\, & \,$K\to\pi\pi\phi\bar\phi'$\,
\\ \hline \hline \small Coefficients &
$\tilde{c}^{\textsc s}$, $\tilde{c}^{\textsc p}$  &
\,$\texttt C^{\textsc v},\texttt C^{\textsc a},\texttt C^{\textsc s},
\texttt C^{\textsc p},\texttt C^{\textsc t},\texttt C^{\textsc t\prime}$\, &
\,$\tilde{c}^{\textsc v},\tilde{c}^{\textsc a},\tilde{c}^{\textsc s},
\tilde{c}^{\textsc p},\texttt C^{\textsc t},\texttt C^{\textsc t\prime}$\, &
$c^{\textsc p}\vphantom{\int_{|^|}^{|^|}}$ & $c^{\textsc v}$, $c^{\textsc s}$ &
$c^{\textsc a}$, $c^{\textsc p}$
\\ \hline
\end{tabular} \bigskip
\end{table}

In Table\,\,\ref{couplings}, we list the contributions of the different constants in
Eqs.\,\,(\ref{Lf}) and\,\,(\ref{Lphi}) to the kaon decays of interest according to
the discussion above.
We remark that in the \,$\texttt f'=\texttt f$\, case for $\texttt f$ having a Majorana nature,
instead of Dirac one,
\,$\overline{\texttt f}\gamma^\eta\texttt f=\overline{\texttt f}\sigma^{\eta\kappa}\texttt f=0$,\,
which causes the $\texttt C_{\texttt{ff}}^{\textsc v,\textsc t,\textsc t\prime}$
and $\tilde{c}_{\texttt{ff}}^{\textsc v}$ parts to disappear.
Moreover, for \,$\phi'=\phi$\, being a real field, rather than complex one,
the $c_{\phi\phi}^{\textsc v,\textsc a}$ terms would be absent.

For later convenience, here we write down the numerical branching fractions of \,$K\to\slashed E$\,
and \,$K\to\pi\pi\slashed E$\, in terms of the contributing coefficients, employing the central
values of the measured kaon lifetimes and meson masses from Ref.\,\cite{Zyla:2020zbs} as well as
\,$a_T^{}=0.658(23)$/GeV\, from lattice QCD work \cite{Baum:2011rm}.
Before doing so, in view of Eqs.\,\,(\ref{GK2ff}) and\,\,(\ref{G''K2pp'ff}) and the fact that
$\texttt f$ and $\texttt f'$ are not detected in the searches, it is appropriate to define
\begin{align} \label{ff'}
{\cal B}(K\to(\pi\pi)\texttt{ff}') \,=\,
\frac{{\cal B}\big(K\to(\pi\pi)\texttt f\bar{\texttt f}{}'\big)
+ {\cal B}\big(K\to(\pi\pi)\texttt f'\bar{\texttt f}\big)}{1+\delta_{\texttt{ff}'}} \,, &
\end{align}
where the factor $1/(1+\delta_{\texttt{ff}'})$ with the Kronecker delta
$\delta_{\texttt{ff}'}$ has been added to prevent double counting when
\,$\texttt f'=\texttt f$.\,
Thus, Eq.\,(\ref{GK2ff}) translates into
\begin{align} \label{BK2ff}
{\cal B}\big(K_L^{}\to\texttt{ff}'\big) & \,=\, 1.45\, \Big(
\big|\tilde{c}_{\texttt{ff}'}^{\textsc s}
- \tilde{c}_{\texttt f'\texttt f}^{\textsc s*}\big|^2
+ \big|\tilde{c}_{\texttt{ff}'}^{\textsc p}
+ \tilde{c}_{\texttt f'\texttt f}^{\textsc p*}\big|^2 \Big)
\frac{10^{14}\rm\,GeV^4}{1+\delta_{\texttt{ff}'}} \,, ~~~ ~~
\nonumber \\ \vphantom{|^{|_|^{|_|^|}}}
{\cal B}\big(K_S^{}\to\texttt{ff}'\big) & \,=\, 2.54\, \Big(
\big|\tilde{c}_{\texttt{ff}'}^{\textsc s}
+ \tilde{c}_{\texttt f'\texttt f}^{\textsc s*}\big|^2
+ \big|\tilde{c}_{\texttt{ff}'}^{\textsc p}
- \tilde{c}_{\texttt f'\texttt f}^{\textsc p*}\big|^2 \Big)
\frac{10^{11}\rm\,GeV^4}{1+\delta_{\texttt{ff}'}}  \,,
\end{align}
and Eq.\,(\ref{G''K2pp'ff}) yields
\begin{subequations} \label{BK2ppff}
\begin{align}
{\cal B}\big(K^-\to\pi^0\pi^-\texttt{ff}'\big) = & ~\Big[
6.28\, \Big( \big|\tilde{c}_{\texttt{ff}'}^{\textsc v}\big|^2
+ \big|\tilde{c}_{\texttt{ff}'}^{\textsc a}\big|^2
+ \big|\tilde{c}_{\texttt f'\texttt f}^{\textsc v}\big|^2
+ \big|\tilde{c}_{\texttt f'\texttt f}^{\textsc a}\big|^2 \Big)
\nonumber \\ & ~+\,
2.01\, \Big( \big|\tilde{c}_{\texttt{ff}'}^{\textsc s}\big|^2
+ \big|\tilde{c}_{\texttt{ff}'}^{\textsc p}\big|^2
+ \big|\tilde{c}_{\texttt f'\texttt f}^{\textsc s}\big|^2
+ \big|\tilde{c}_{\texttt f'\texttt f}^{\textsc p}\big|^2 \Big)
\nonumber \\ & ~+\,
6.59\, \Big( \big|\texttt C_{\texttt{ff}'}^{\textsc t}\big|^2
+ \big|\texttt C_{\texttt{ff}'}^{\textsc t\prime}\big|^2
+ \big|\texttt C_{\texttt f'\texttt f}^{\textsc t}\big|^2
+ \big|\texttt C_{\texttt f'\texttt f}^{\textsc t\prime}\big|^2 \Big) \Big]
\frac{10^5 \rm\,GeV^4}{1+\delta_{\texttt{ff}'}} \,, & ~~~
\end{align}
\vspace{-3ex}
\begin{align}
{\cal B}\big(K_L^{}\to\pi^+\pi^-\texttt{ff}'\big) = & ~\Big\{
13.4\, \Big[ \big|\tilde{c}_{\texttt{ff}'}^{\textsc v}
- \tilde{c}_{\texttt f'\texttt f}^{\textsc v*}\big|^2
+ \big|\tilde{c}_{\texttt{ff}'}^{\textsc a}
- \tilde{c}_{\texttt f'\texttt f}^{\textsc a*}\big|^2 \Big]
+ 67.7\, \Big[ \big|\tilde{c}_{\texttt{ff}'}^{\textsc v}
+ \tilde{c}_{\texttt f'\texttt f}^{\textsc v*}\big|^2
+ \big|\tilde{c}_{\texttt{ff}'}^{\textsc a}
+ \tilde{c}_{\texttt f'\texttt f}^{\textsc a*}\big|^2 \Big]
\nonumber \\ & ~+\,
4.31\,\Big[ \big|\tilde{c}_{\texttt{ff}'}^{\textsc s}
+ \tilde{c}_{\texttt f'\texttt f}^{\textsc s*}\big|^2
+ \big|\tilde{c}_{\texttt{ff}'}^{\textsc p}
- \tilde{c}_{\texttt f'\texttt f}^{\textsc p*}\big|^2 \Big]
+ 118\, \Big[ \big|\tilde{c}_{\texttt{ff}'}^{\textsc s}
- \tilde{c}_{\texttt f'\texttt f}^{\textsc s*}\big|^2
+ \big|\tilde{c}_{\texttt{ff}'}^{\textsc p}
+ \tilde{c}_{\texttt f'\texttt f}^{\textsc p*}\big|^2 \Big]
\nonumber \\ & ~+\, 14.3\,
\Big[ \big|{\tt C}_{\texttt{ff}'}^{\textsc t}+{\tt C}_{\texttt f'\texttt f}^{\textsc t*}\big|^2
+ \big|{\tt C}_{\texttt{ff}'}^{\textsc t\prime}-{\tt C}_{\texttt f'\texttt f}^{\textsc t\prime*}|^2
\Big]
\nonumber \\ & ~+\, 0.50\,
\Big[ \big|{\tt C}_{\texttt{ff}'}^{\textsc t}-{\tt C}_{\texttt f'\texttt f}^{\textsc t\prime*}
\big|^2 + \big|{\tt C}_{\texttt{ff}'}^{\textsc t}+{\tt C}_{\texttt f'\texttt f}^{\textsc t\prime*}
\big|^2 \Big] \Big\} \frac{10^5 \rm\,GeV^4}{1+\delta_{\texttt{ff}'}} \,,
\end{align}
\vspace{-3ex}
\begin{align}
{\cal B}\big(K_L^{}\to\pi^0\pi^0\texttt{ff}'\big) = & ~\Big\{
42.4\, \Big[ \big|\tilde{c}_{\texttt{ff}'}^{\textsc v}
+ \tilde{c}_{\texttt f'\texttt f}^{\textsc v*}\big|^2
+ \big|\tilde{c}_{\texttt{ff}'}^{\textsc a}
+ \tilde{c}_{\texttt f'\texttt f}^{\textsc a*}\big|^2 \Big]
\nonumber \\ & ~+\,
80.2\, \Big[ \big|\tilde{c}_{\texttt{ff}'}^{\textsc s}
- \tilde{c}_{\texttt f'\texttt f}^{\textsc s*}\big|^2
+ \big|\tilde{c}_{\texttt{ff}'}^{\textsc p}
+ \tilde{c}_{\texttt f'\texttt f}^{\textsc p*}\big|^2 \Big]
\nonumber \\ & ~+\,
0.34\, \Big[ \big| {\tt C}_{\texttt{ff}'}^{\textsc t} - {\tt C}_{\texttt f'\texttt f}^{\textsc t*}
\big|^2 + \big| {\tt C}_{\texttt{ff}'}^{\textsc t\prime}
+ {\tt C}_{\texttt f'\texttt f}^{\textsc t\prime*} \big|^2
\Big] \Big\} \frac{10^5 \rm\,GeV^4}{1+\delta_{\texttt{ff}'}} \,. &
\end{align}
\end{subequations}
In the invisible scalar case, analogously to Eq.\,(\ref{ff'}), we express
\begin{align}
{\cal B}\big(K\to(\pi\pi)\phi\phi'\big) & \,=\, \frac{{\cal B}\big(K\to(\pi\pi)\phi\bar\phi'\big)
+ {\cal B}\big(K\to(\pi\pi)\phi'\bar\phi\big)}{1+\delta_{\phi\phi'}} \,. &
\end{align}
From Eqs.\,\,(\ref{GK2ss}) and (\ref{G''K2pp'ss}) we then have
\begin{align} \label{BK2ss}
{\cal B}\big(K_L^{}\to\phi\phi'\big) & =\, 2.93 \times 10^{14} \rm\; GeV^2\, \frac{
\big| c_{\phi\phi'}^{\textsc p} - c_{\phi'\phi}^{\textsc p*} \big|\raisebox{2pt}{$^2$}}
{1+\delta_{\phi\phi'}} \,,
\nonumber \\
{\cal B}\big(K_S^{}\to\phi\phi'\big) & =\, 5.14 \times 10^{11} \rm\; GeV^2\, \frac{
\big| c_{\phi\phi'}^{\textsc p} + c_{\phi'\phi}^{\textsc p*} \big|\raisebox{2pt}{$^2$}}
{1+\delta_{\phi\phi'}} \, &
\end{align}
and
\begin{align} \label{BK2ppss}
{\cal B}\big(K^-\to\pi^0\pi^-\phi\phi'\big) & = \Big[ 0.0157\, \Big(
\big|c_{\phi\phi'}^{\textsc a}\big|^2 + \big|c_{\phi'\phi}^{\textsc a}\big|^2 \Big)
+ 1.38\, \Big( \big|c_{\phi\phi'}^{\textsc p}\big|^2 + \big|c_{\phi'\phi}^{\textsc p}\big|^2 \Big)
{\rm GeV}^{-2} \Big] \frac{10^7 \rm\,GeV^4}{1+\delta_{\phi\phi'}} \,,
\nonumber \\ {\cal B}\big(K_L^{}\to\pi^+\pi^-\phi\phi'\big) & = \! \begin{array}[t]{l} \displaystyle
\Big( 0.0334\, \big|c_{\phi'\phi}^{\textsc a*}-c_{\phi\phi'}^{\textsc a}\big|^2
+ 2.94\, \big|c_{\phi'\phi}^{\textsc p*}+c_{\phi\phi'}^{\textsc p}\big|^2 {\rm\,GeV}^{-2}
\\ \displaystyle \,+~
0.169\, \big|c_{\phi'\phi}^{\textsc a*}+c_{\phi\phi'}^{\textsc a}\big|^2
+ 51.3\, \big|c_{\phi'\phi}^{\textsc p*}-c_{\phi\phi'}^{\textsc p}\big|^2 {\rm\,GeV}^{-2} \Big)
\frac{10^7 \rm\,GeV^4}{1+\delta_{\phi\phi'}} \,, \end{array}
\nonumber \\ \vphantom{|^{|^{|_|^|}}}
{\cal B}\big(K_L^{}\to\pi^0\pi^0\phi\phi'\big) & = \Big(
0.106\, \big|c_{\phi'\phi}^{\textsc a*}+c_{\phi\phi'}^{\textsc a}\big|^2
+ 32.1\, \big|c_{\phi'\phi}^{\textsc p*}-c_{\phi\phi'}^{\textsc p}\big|^2 {\rm\,GeV}^{-2} \Big)
\frac{10^7 \rm\,GeV^4}{1+\delta_{\phi\phi'}} \,, &
\end{align}
respectively.

\section{SM predictions and empirical information\label{smB}}

As mentioned earlier, the latest NA62 measurement on \,$K^+\to\pi^+\nu\bar\nu$\, has turned up
evidence for it that is fully consistent with the SM expectation~\cite{na62new}.
In view of Table\,\,\ref{couplings}, this implies that the couplings
$\texttt C_{\texttt f}^{\textsc v,\textsc a,\textsc s,\textsc p,\textsc t,\textsc t\prime}$
and $c_\phi^{\textsc v,\textsc s}$ originating from possible NP cannot by sizable
anymore.\footnote{A preliminary report from KOTO \cite{kotonew} has revealed that its most recent
data contain a couple of \,$K_L\to\pi^0\nu\bar\nu$\, events suggesting an anomalously high rate,
which still needs confirmation from further measurements.
If this anomaly persists in the future, it may be due to NP, as discussed in $e.g.$
\cite{Egana-Ugrinovic:2019wzj,Dev:2019hho,He:2020jly} and the references therein, but its
effects would not be large enough to modify our conclusions for $K\to\pi\pi\slashed E$.}
To explore how much the other coefficients shown in Table\,\,\ref{couplings} may be affected
by NP to amplify the $K\to\slashed E$ and $K\to\pi\pi\slashed E$ rates with respect to their
SM values, we need to know the latter.

In the SM, our processes of interest arise at short distance from effective \,$ds\nu_l\bar\nu_l$\,
interactions, with \,$l=e,\mu,\tau$,\, described by \cite{Buchalla:1995vs}
\begin{align} \label{Lsm}
{\cal L}_{sd\nu\nu}^{\textsc{sm}} & \,=\, \frac{-\alpha_{\rm e}^{}G_{\rm F}^{}}{\sqrt8\,\pi
\sin^2\!\theta_{\textsc w}}\, \raisebox{2pt}{\footnotesize$\displaystyle\sum_{l=e,\mu,\tau}$}
\big( V_{td}^*V_{ts}^{} X_t^{}+V_{cd}^*V_{cs}^{} X_c^l\big)\, \overline{d}\gamma^\eta
(1-\gamma_5^{})s~ \overline{\nu_l^{}}\gamma_\eta^{}(1-\gamma_5^{})\nu_l^{}
\;+\; {\rm H.c.} \,, &
\end{align}
where \,$\alpha_{\rm e}^{}=1/128$,\, $G_{\rm F}$ is the Fermi constant,
\,$\sin^2\!\theta_{\textsc w}=0.231$, $V_{qq'}$ are Cabibbo-Kobayashi-Maskawa (CKM) matrix
elements, \,$X_t=1.481$\, from $t$-quark loops, and \,$X_c^e=X_c^\mu\simeq1.2\times10^{-3}$ and
\,$X_c^\tau\simeq8\times10^{-4}$ are $c$-quark contributions \cite{Buchalla:1995vs}.
Applying the notation of Eq.\,(\ref{Lf}) to ${\cal L}_{sd\nu\nu}^{\textsc{sm}}$, we then have
\,$\texttt C_{\nu_l\nu_l}^{\textsc v}=-\texttt C_{\nu_l\nu_l}^{\textsc a}
= -\tilde{c}_{\nu_l\nu_l}^{\textsc v}
= \tilde{c}_{\nu_l\nu_l}^{\textsc a}=\alpha_{\rm e}^{}G_{\rm F}
\big(\lambda_t^{}X_t^{}+ \lambda_c^{}X_c^l\big)/\big(\sqrt8\,\pi\sin^2\!\theta_{\textsc w}\big)
\sim -(3+0.9i)\times10^{-11}/\rm GeV^2$\,
and
\,$\texttt C_{\nu_l\nu_l}^{\textsc s,\textsc p,\textsc t,\textsc t\prime} =
\tilde{c}_{\nu_l\nu_l}^{\textsc s,\textsc p} = 0$.\,

Accordingly, in light of Eqs.\,\,(\ref{GK2ff}) and (\ref{SK2ff}) we see that
\,${\mathcal B}(K_{L,S}\to\nu\bar\nu)_{\textsc{sm}}^{}=0$,\, given that the neutrinos are
massless in the SM.
However, supplementing it with nonzero neutrino masses and taking their biggest one from
the direct limit \,$m_{\nu_\tau}^{\rm exp}<18.2$\,MeV \cite{Zyla:2020zbs} would instead lead
to the maximal values~\cite{Tandean:2019tkm}
${\cal B}(K_L\to\nu\bar\nu)_{\textsc{sm}}^{}\simeq1\times10^{-10}$ and
\,${\cal B}(K_S\to\nu\bar\nu)_{\textsc{sm}}^{}\simeq2\times10^{-14}$.\,

As for the four-body channels, employing Eq.\,(\ref{BK2ppff}) we get
\,${\mathcal B}(K^-\to\pi^0\pi^-\nu\bar\nu)_{\textsc{sm}}^{} \sim 4\times10^{-15}$ and
\,${\mathcal B}(K_L\to(\pi^+\pi^-,\pi^0\pi^0)\nu\bar\nu)_{\textsc{sm}}^{} \sim
(8,5)\times10^{-14}$.\,
These are in rough agreement with more refined evaluations in
the literature~\cite{Geng:1994cw,Littenberg:1995zy,Chiang:2000bg,Geng:1996kd}:
\begin{align} \label{smK2ppnn}
{\mathcal B}(K^-\to\pi^0\pi^-\nu\bar\nu)_{\textsc{sm}}^{} & \,\sim\, 1.2\times10^{-14} \,, &
{\mathcal B}(K_L\to\pi^+\pi^-\nu\bar\nu)_{\textsc{sm}}^{} & \,\sim\, 2.8\times10^{-13} \,, ~~~
\nonumber \\
{\mathcal B}(K_L\to\pi^0\pi^0\nu\bar\nu)_{\textsc{sm}}^{} & \,\sim\, 1.5\times10^{-13} \,.
\end{align}
with the latest CKM matrix elements~\cite{Zyla:2020zbs}.
The estimates for \,$K_S\to\pi\pi\nu\bar\nu$\, are about three orders of magnitude less than
their $K_L$ counterparts.
The two sets of $K^-$ and $K_L$ numbers above indicate the level of uncertainties in our
\,$K\to\pi\pi\slashed E$\, predictions in the next section.

On the experimental side, merely two of these modes have been looked for~\cite{Zyla:2020zbs},
with negative outcomes which led to the limits~\cite{Adler:2000ic,E391a:2011aa}
\begin{align} \label{xBK2ppinv}
{\mathcal B}(K^-\to\pi^0\pi^-\nu\bar\nu)_{\rm exp}^{} & \,<\, 4.3\times10^{-5} \,, &
{\mathcal B}(K_L\to\pi^0\pi^0\nu\bar\nu)_{\rm exp}^{} & \,<\, 8.1\times10^{-7} &
\end{align}
both at 90\%~CL.
These exceed the corresponding SM numbers in Eq.\,(\ref{smK2ppnn}) by several orders of magnitude.
As regards \,$K_{L,S}\to\slashed E$,\, there have been no direct searches for them yet.
Nevertheless, from the existing data \cite{Zyla:2020zbs} on the visible decay channels of
$K_{L,S}$ one can obtain indirect upper bounds on their invisible branching
fractions~\cite{Gninenko:2014sxa}.
Thus, one can infer~\cite{Su:2020xwt}
\begin{align} \label{xBK2inv}
{\mathcal B}(K_L\to\slashed E) & \,<\, 1.8\times10^{-3} \,, &
{\mathcal B}(K_S\to\slashed E) & \,<\, 7.1\times10^{-4} ~~~ ~~
\end{align}
at the 2$\sigma$ level, which are far away from the aforesaid
${\cal B}(K_{L,S}\to\nu\bar\nu)_{\textsc{sm}}^{}$ values.
Comparing Eqs.\,(\ref{xBK2ppinv})-(\ref{xBK2inv}) with Eqs.\,(\ref{BK2ff})-(\ref{BK2ppss}),
as well as Table\,\,\ref{couplings}, we conclude that currently there remains potentially
plenty of room for NP to boost the rates of these decays via
$\tilde{c}^{\textsc v,\textsc a,\textsc s,\textsc p}$
and $c^{\textsc a,\textsc p}$.

\section{NP expectations and conclusions\label{bsm}}

Based on the considerations made in the previous section, we hereafter entertain the possibility
that, among the couplings listed in the table, NP manifests itself exclusively via those
belonging to operators with parity-odd $ds$ bilinears, namely
$\tilde{c}^{\textsc v,\textsc a,\textsc s,\textsc p}$ in Eq.\,(\ref{Lf}) or
$c^{\textsc a,\textsc p}$ in Eq.\,(\ref{Lphi}), and demand that they fulfill the conditions
\begin{align} \label{npK2e}
{\mathcal B}(K_L\to\slashed E)_{\textsc{np}}^{} & \,<\, 1.8\times10^{-3} \,, &
{\mathcal B}(K_S\to\slashed E)_{\textsc{np}}^{} & \,<\, 7.1\times10^{-4} \,, ~~~ ~~
\\ \label{npK2ppe}
{\mathcal B}(K^-\to\pi^0\pi^-\slashed E)_{\textsc{np}}^{} & \,<\, 4.0\times10^{-5} \,, &
{\mathcal B}(K_L\to\pi^0\pi^0\slashed E)_{\textsc{np}}^{} & \,<\, 8.0\times10^{-7} \,.
\end{align}
These $ds\texttt{ff}'$, or $ds\phi\phi'$, interactions additionally contribute to the mixing of
neutral kaons via one-loop diagrams with \texttt f and \texttt f$'$, or $\phi$ and $\phi'$,
being in the loops and must therefore be compatible with its data.
One can see that the resulting pertinent operators are of the form
\,$\overline d(\gamma^\kappa)\gamma_5^{}s\,\overline d(\gamma_\kappa^{})\gamma_5^{}s$\,
and have coefficients proportional to linear combinations of
\,$\tilde c_{\texttt{ff}'}^x\tilde c_{\texttt f'\texttt f}^x$\, with
\,$x=\textsc v,\textsc a,\textsc s,\textsc p$,\, or
\,$c_{\phi\phi'}^xc_{\phi'\phi}^x$\, with \,$x=\textsc a,\textsc p$.\,
As a consequence, these products can evade the kaon-mixing restrictions, which are stringent,
if one of $\tilde c_{\texttt{ff}'}^x$ and $\tilde c_{\texttt f'\texttt f}^x$ for
\,$\texttt f'\neq\texttt f$,\, or one of $c_{\phi\phi'}^x$ and $c_{\phi'\phi}^x$ for
\,$\phi'\neq\phi$,\, either vanishes or is sufficiently smaller than the other.
To illustrate the ramifications that may arise for the various \,$K\to\pi\pi\slashed E$\, modes
if NP occurs in these couplings, we can look at several simple examples.

If it solely affects $\tilde{c}_{\texttt{ff}'}^{\textsc s,\textsc p}$ with
\,$\texttt f'\neq\texttt f$,\, then $\tilde{c}_{\texttt f'\texttt f}^{\textsc s,\textsc p}$ are
absent, and so the kaon-mixing constraints are avoided.
In this case, comparing Eqs.\,\,(\ref{BK2ff})-(\ref{BK2ppff}) to (\ref{npK2e})-(\ref{npK2ppe}),
we learn that the \,$K_L\to\slashed E$\, requirement is the most significant and translates into
\,$|\tilde{c}_{\texttt{ff}'}^{\textsc s}|^2 + |\tilde{c}_{\texttt{ff}'}^{\textsc p}|^2
< 1.2\times10^{-17}{\rm\;GeV}^{-4}$.\,
Combining it with the branching fractions, we arrive at the maximal values
\begin{align} \label{new1BK2ppff}
{\cal B}\big(K^-\to\pi^0\pi^-\texttt{ff}'\big) & \,<\, 2.5\times10^{-12} \,, &
{\cal B}\big(K_L^{}\to\pi^+\pi^-\texttt{ff}'\big) & \,<\, 1.5\times10^{-10} \,, ~~~
\nonumber \\
{\cal B}\big(K_L^{}\to\pi^0\pi^0\texttt{ff}'\big) & \,<\, 1.0\times10^{-10} \,, &
{\cal B}\big(K_S^{}\to\pi^+\pi^-\texttt{ff}'\big) & \,<\, 2.7\times10^{-13} \,,
\nonumber \\
{\cal B}\big(K_S^{}\to\pi^0\pi^0\texttt{ff}'\big) & \,<\, 1.7\times10^{-13} \,.
\end{align}
Interchanging the roles of $\tilde{c}_{\texttt{ff}'}^{\textsc s,\textsc p}$ and
$\tilde{c}_{\texttt f'\texttt f}^{\textsc s,\textsc p}$ would not alter these numbers.
They are considerably higher than the corresponding SM expectations quoted earlier but
might not be high enough to be empirically testable any time soon.

If only $\tilde{c}_{\texttt{ff}'}^{\textsc v,\textsc a}$ with \,$\texttt f'\neq\texttt f$\, are
influenced by NP, hence \,$\tilde{c}_{\texttt f'\texttt f}^{\textsc v,\textsc a}=0$,\,
it is clear from Eq.\,(\ref{BK2ff}) that Eq.\,(\ref{npK2e}) is no longer
relevant but Eq.\,(\ref{npK2ppe}) still matters, with the \,$K_L\to\pi^0\pi^0\slashed E$\,
restraint being the stronger and yielding
\,$|\tilde{c}_{\texttt{ff}'}^{\textsc v}|^2
+ |\tilde{c}_{\texttt{ff}'}^{\textsc a}|^2 < 1.9\times10^{-13}{\rm\;GeV}^{-4}$.\,
With this, we obtain
\begin{align} \label{new2BK2ppff}
{\cal B}\big(K^-\to\pi^0\pi^-\texttt{ff}'\big) & \,<\, 1.2\times10^{-7} \,, &
{\cal B}\big(K_L^{}\to\pi^+\pi^-\texttt{ff}'\big) & \,<\, 1.5\times10^{-6} \,,
\nonumber \\
{\cal B}\big(K_L^{}\to\pi^0\pi^0\texttt{ff}'\big) & \,<\, 8.0\times10^{-7} \,, &
{\cal B}\big(K_S^{}\to\pi^+\pi^-\texttt{ff}'\big) & \,<\, 2.7\times10^{-9}  \,,
\nonumber \\
{\cal B}\big(K_S^{}\to\pi^0\pi^0\texttt{ff}'\big) & \,<\, 1.4\times10^{-9} \,,
\end{align}
which greatly exceed their counterparts in Eq.\,(\ref{new1BK2ppff}) and some of which may be
within the reach of ongoing or upcoming kaon factories.
We would achieve the same results with $\tilde{c}_{\texttt f'\texttt f}^{\textsc v,\textsc a}$\,
alone instead.
It is worth pointing out that this kind of possibility can be realized in a scenario involving
scalar leptoquarks plus light sterile neutrinos acting as the invisibles~\cite{Su:2019tjn}.
Furthermore, the model can also generate substantial enhancement in the rates of
the aforementioned FCNC hyperon decays with missing energy~\cite{Su:2019tjn}, which are
potentially detectable in the BESIII experiment~\cite{Li:2016tlt,Ablikim:2019hff}.

If now NP enters exclusively through $c_{\phi\phi'}^{\textsc p}$ with \,$\phi'\neq\phi$,\,
implying \,$c_{\phi'\phi}^{\textsc p}=0$,\, then a comparison of
Eqs.\,\,(\ref{BK2ss})-(\ref{BK2ppss}) and (\ref{npK2e})-(\ref{npK2ppe}) reveals that
the \,$K_L\to\slashed E$\, requisite in Eq.\,(\ref{npK2e}) is the most important, from which we get
\,$\big|c_{\phi\phi'}^{\textsc p}\big|\raisebox{1pt}{$^2$}<6.1\times10^{-18}{\rm\;GeV}^{-2}$.\,
This translates into
\begin{align} \label{new1BK2ppss}
{\cal B}\big(K^-\to\pi^0\pi^-\phi\bar\phi\big) & \,<\, 8.5\times10^{-11} \,, &
{\cal B}\big(K_L^{}\to\pi^+\pi^-\phi\bar\phi\big) & \,<\, 3.3\times10^{-9} \,,
\nonumber \\
{\cal B}\big(K_L^{}\to\pi^0\pi^0\phi\bar\phi\big) & \,<\, 2.0\times10^{-9} \, &
{\cal B}\big(K_S^{}\to\pi^+\pi^-\phi\bar\phi\big) & \,<\, 5.8\times10^{-12} \,,
\nonumber \\
{\cal B}\big(K_S^{}\to\pi^0\pi^0\phi\bar\phi\big) & \,<\, 3.5\times10^{-12}
\end{align}
which are larger than the corresponding values in Eq.\,(\ref{new1BK2ppff}) by roughly
an order of magnitude.
In contrast, if NP solely impacts $c_{\phi\phi'}^{\textsc a}$ with \,$\phi'\neq\phi$,\,
hence \,$c_{\phi'\phi}^{\textsc a}=0$,\, the situation turns out to be analogous to that
reflected by Eq.\,(\ref{new2BK2ppff}).
More explicitly, in view of Eqs.\,\,(\ref{BK2ppss}) and (\ref{npK2ppe}), from the $K_L$
condition in the latter we extract
\,$\big|c_{\phi\phi'}^{\textsc a}\big|\raisebox{1pt}{$^2$}<7.5\times10^{-13}{\rm\;GeV}^{-4}$,\,
which leads to
\begin{align} \label{new2BK2ppss}
{\cal B}\big(K^-\to\pi^0\pi^-\phi\bar\phi\big) & \,<\, 1.2\times10^{-7} \,, &
{\cal B}\big(K_L^{}\to\pi^+\pi^-\phi\bar\phi\big) & \,<\, 1.5\times10^{-6} \,,
\nonumber \\
{\cal B}\big(K_L^{}\to\pi^0\pi^0\phi\bar\phi\big) & \,<\, 8.0\times10^{-7} \, &
{\cal B}\big(K_S^{}\to\pi^+\pi^-\phi\bar\phi\big) & \,<\, 2.7\times10^{-9} \,,
\nonumber \\
{\cal B}\big(K_S^{}\to\pi^0\pi^0\phi\bar\phi\big) & \,<\, 1.4\times10^{-9} \,.
\end{align}
These numbers are identical to those in Eq.\,(\ref{new2BK2ppff}).

\begin{table}[t]
\caption{The maximal branching fractions of \,$K\to\pi\pi\slashed E$\, due to NP being present
in one or more of the coefficients
$\tilde{c}_{\texttt{ff}',\texttt f'\texttt f}^{\textsc v,\textsc a,\textsc s,\textsc p}$ and
$c_{\phi\phi',\phi'\phi}^{\textsc a,\textsc p}$ as specified in the examples given in the main
text.\label{BK}}
\begin{tabular}{|c||c|c|c|c|c|} \hline
\multirow{2}{*}{$\begin{array}{c}\rm Contributing \vspace{-3pt} \\ \rm coefficients \end{array}$} &
\multicolumn{5}{c|}{Decay modes} \\ \cline{2-6} & \,$K^-\to\pi^0\pi^-\slashed E$\, &
\,$K_L\to\pi^+\pi^-\slashed E$\, & \,$K_L\to\pi^0\pi^0\slashed E^{\vphantom{|^|}}$\, &
\,$K_S\to\pi^+\pi^-\slashed E$\, & \,$K_S\to\pi^0\pi^0\slashed E$\,
\\ \hline \hline
$\begin{array}{c}\tilde c_{\texttt{ff}'_{\vphantom{o}}}^{\textsc s,\textsc p} {\rm\,~or~\,}
\tilde c_{\texttt f'\texttt f}^{\textsc s,\textsc p^{\vphantom{|^|}}} \end{array}$ &
$2.5\times10^{-12}$ & $1.5\times10^{-10}$ & $1.0\times10^{-10}$ & $2.7\times10^{-13}$ &
$1.7\times10^{-13}$
\\ \hline
$\begin{array}{c} \tilde c_{\texttt{ff}'_{\vphantom{o}}}^{\textsc v,\textsc a} {\rm\,~or~\,}
\tilde c_{\texttt f'\texttt f}^{\textsc v,\textsc a^{\vphantom{|^|}}} \end{array}$ &
$1.2\times10^{-7}$ & $1.5\times10^{-6}$ & $8.0\times10^{-7}$ & $2.7\times10^{-9}$ &
$1.4\times10^{-9}$
\\ \hline
$c_{\phi\phi'_{\vphantom{0}}}^{\textsc p^{\vphantom{|^|}}}$ {\rm\,or\,} $c_{\phi'\phi}^{\textsc p}$
& $8.5\times10^{-11}$ & $3.3\times10^{-9}$ & $2.0\times10^{-9}$ & $5.8\times10^{-12}$ &
$3.5\times10^{-12}$
\\ \hline
$c_{\phi\phi'_{\vphantom{0}}}^{\textsc a^{\vphantom{|^|}}}$ {\rm\,or\,}
$c_{\phi'\phi}^{\textsc a}$ & $1.2\times10^{-7}$ & $1.5\times10^{-6}$ & $8.0\times10^{-7}$ &
$2.7\times10^{-9}$ & $1.4\times10^{-9}$
\\ \hline
\end{tabular} \bigskip
\end{table}

In Table\,\,\ref{BK} we collect our findings in Eqs.\,(\ref{new1BK2ppff})-(\ref{new2BK2ppss})
and the associated coefficients.
We note that, as alluded to in Sec.\,\ref{intro} and discussed in
Refs.\,\cite{Tandean:2019tkm,Su:2019tjn,Li:2019cbk}, in the cases seen in this table with
high branching fractions the corresponding predictions for their hyperon counterparts are
magnified in like manner and might therefore be within the sensitivity ranges of searches in
the near future.

To conclude, motivated by the latest NA62 measurement on \,$K^+\to\pi^+\nu\bar\nu$, which is
in good agreement with the SM and consequently implies stringent constraints on NP which
might be hiding in \,$K\to\pi\slashed E$,\, we have explored how other types of FCNC kaon decays
with missing energy might shed additional light on potential NP in the underlying
\,$s\to d\slashed E$\, transition.
Focusing on scenarios in which the missing energy is carried away by a pair of invisible new
particles of spin 1/2 or 0, we have argued that there are four-particle operators
contributing to \,$s\to d\slashed E$\, which are not restricted by \,$K\to\pi\slashed E$\,
and accordingly could still significantly affect \,$K\to\slashed E$ and $K\to\pi\pi\slashed E$,\,
on which the empirical details are currently meager.
We have demonstrated especially that the branching fractions of \,$K\to\pi\pi\slashed E$\,
could yet be amplified far beyond their SM expectations, to levels which might be within
the reach of ongoing experiments, specifically KOTO and NA62,
or upcoming ones such as KLEVER~\cite{Ambrosino:2019qvz}.
Our results, which are illustrated with the instances summarized in Table\,\,\ref{BK},
will hopefully help stimulate new quests for these decays as NP probes.
Last but not least, we have pointed out that similar kinds of enhancement would occur in
the hyperon sector, which may be detectable by BESIII or future charm-tau
factories~\cite{Luo:2018njj,Barnyakov:2020vob}, and thus it could offer complementary
NP tests.

\acknowledgments
This work was supported in part by National Center for Theoretical Sciences and
MoST (Grants No. MOST-106-2112-M-002-003-MY3 and No. MoST-107-2119-M-007-013-MY3).


\begin{thebibliography}{0}

\bibitem{Littenberg:1993qv}
  L.~Littenberg and G.~Valencia,
``Rare and radiative kaon decays,''
Ann.\ Rev.\ Nucl.\ Part.\ Sci.\ {\bf 43}, 729 (1993) 
  [hep-ph/9303225].

\bibitem{Buchalla:1995vs}
G.~Buchalla, A.J.~Buras, and M.E.~Lautenbacher,
``Weak decays beyond leading logarithms,''
  Rev.\ Mod.\ Phys.\  {\bf 68}, 1125 (1996) 
  [hep-ph/9512380].

\bibitem{Geng:1994cw}
  C.Q.~Geng, I.J.~Hsu, and Y.C.~Lin,
``CP conserving and violating contributions to $K_L\to\pi^+\pi^-\nu\bar\nu$'',
  Phys.\ Rev.\ D {\bf 50}, 5744 (1994) 
  [hep-ph/9406313].

\bibitem{Geng:1996kd}
  C.Q.~Geng, I.J.~Hsu, and Y.C.~Lin,
``Study of long distance contributions to $K\to n\pi\nu\bar\nu$'',
  Phys.\ Rev.\ D {\bf 54}, 877 (1996) 
  [hep-ph/9604228].

\bibitem{Cirigliano:2011ny}
  V.~Cirigliano, G.~Ecker, H.~Neufeld, A.~Pich, and J.~Portoles,
``Kaon Decays in the Standard Model,''
  Rev.\ Mod.\ Phys.\ {\bf 84}, 399 (2012) 
  [arXiv:1107.6001 [hep-ph]].

\bibitem{Tandean:2019tkm}
  J.~Tandean,
``Rare hyperon decays with missing energy,''
  JHEP {\bf 1904}, 104 (2019) 
  [arXiv:1901.10447 [hep-ph]].

\bibitem{Su:2019tjn}
  J.Y.~Su and J.~Tandean,
``Exploring leptoquark effects in hyperon and kaon decays with missing energy,''
 Phys.\ Rev.\ D {\bf 102}, no. 7, 075032 (2020) 
  [arXiv:1912.13507 [hep-ph]].

\bibitem{Li:2019cbk}
  G.~Li, J.Y.~Su, and J.~Tandean,
``Flavor-changing hyperon decays with light invisible bosons,''
  Phys.\ Rev.\ D {\bf 100}, no. 7, 075003 (2019) 
  [arXiv:1905.08759 [hep-ph]].

\bibitem{Artamonov:2008qb}
  A.V.~Artamonov {\it et al.} [E949 Collaboration],
``New measurement of the $K^{+} \to \pi^{+} \nu \bar{\nu}$ branching ratio'',
  Phys.\ Rev.\ Lett.\  {\bf 101}, 191802 (2008) 
  [arXiv:0808.2459 [hep-ex]].

\bibitem{Ahn:2018mvc}
  J.K.~Ahn {\it et al.} [KOTO Collaboration],
``Search for the $K_L\to\pi^0\nu\overline{\nu}$ and $K_L\to\pi^0X^0$ decays at the J-PARC
KOTO experiment,''
  Phys.\ Rev.\ Lett.\  {\bf 122}, no. 2, 021802 (2019) 
  [arXiv:1810.09655 [hep-ex]].

\bibitem{CortinaGil:2018fkc}
  E.~Cortina Gil {\it et al.} [NA62 Collaboration],
``First search for $K^+\rightarrow\pi^+\nu\bar{\nu}$ using the decay-in-flight technique,''
  Phys.\ Lett.\ B {\bf 791}, 156 (2019) 
  [arXiv:1811.08508 [hep-ex]].

\bibitem{CortinaGil:2020vlo}
  E.~Cortina Gil {\it et al.} [NA62 Collaboration],
``An investigation of the very rare $ {K}^{+}\to {\pi}^{+}\nu \overline{\nu} $ decay,''
  JHEP {\bf 2011}, 042 (2020) 
  [arXiv:2007.08218 [hep-ex]].

\bibitem{Buras:2015qea}
  A.J.~Buras, D.~Buttazzo, J.~Girrbach-Noe, and R.~Knegjens,
``$ {K}^{+}\to {\pi}^{+}\nu \overline{\nu} $ and $ {K}_L\to {\pi}^0\nu \overline{\nu} $ in
the Standard Model: status and perspectives,''
  JHEP {\bf 1511}, 033 (2015) 
  [arXiv:1503.02693 [hep-ph]].

\bibitem{na62new}
R. Marchevski, ``Evidence for the decay $K^+\to\pi^+\nu\bar\nu$ from the NA62 experiment at CERN'',
talk at the 40th International Conference on High Energy Physics (ICHEP 2020), Prague, Czech
Republic, 30 July--5 August 2020.

\bibitem{Badin:2010uh}
  A.~Badin and A.A.~Petrov,
``Searching for light Dark Matter in heavy meson decays,''
  Phys.\ Rev.\ D {\bf 82}, 034005 (2010) 
  [arXiv:1005.1277 [hep-ph]].

\bibitem{Kamenik:2011vy}
  J.F.~Kamenik and C.~Smith,
``FCNC portals to the dark sector,''
  JHEP {\bf 1203}, 090 (2012) 
  [arXiv:1111.6402 [hep-ph]].

\bibitem{Zyla:2020zbs}
  P.A.~Zyla {\it et al.} [Particle Data Group],
``Review of Particle Physics,''
  PTEP {\bf 2020}, no. 8, 083C01 (2020). 

\bibitem{Fabbrichesi:2017vma}
  M.~Fabbrichesi, E.~Gabrielli, and B.~Mele,
``Hunting down massless dark photons in kaon physics'',
  Phys.\ Rev.\ Lett.\  {\bf 119}, no. 3, 031801 (2017) 
  [arXiv:1705.03470 [hep-ph]].

\bibitem{Su:2019ipw}
  J.Y.~Su and J.~Tandean,
``Searching for dark photons in hyperon decays,''
  Phys.\ Rev.\ D {\bf 101}, no. 3, 035044 (2020) 
  [arXiv:1911.13301 [hep-ph]].

\bibitem{Su:2020xwt}
  J.Y.~Su and J.~Tandean,
``Kaon decays shedding light on massless dark photons,''
Eur.\ Phys.\ J.\ C {\bf 80}, 824 (2020) 
  [arXiv:2006.05985 [hep-ph]].

\bibitem{MartinCamalich:2020dfe}
  J.~Martin Camalich, M.~Pospelov, P.N.H.~Vuong, R.~Ziegler, and J.~Zupan,
``Quark Flavor Phenomenology of the QCD Axion,''
  Phys.\ Rev.\ D {\bf 102}, no. 1, 015023 (2020) 
  [arXiv:2002.04623 [hep-ph]].

\bibitem{He:2005we}
  X.G.~He, J.~Tandean, and G.~Valencia,
``Implications of a new particle from the hyperCP data on $\Sigma^+\to p\mu^+\mu-$'',
  Phys.\ Lett.\ B {\bf 631}, 100 (2005) 
  [hep-ph/0509041].

\bibitem{Baum:2011rm}
  I.~Baum, V.~Lubicz, G.~Martinelli, L.~Orifici, and S.~Simula,
``Matrix elements of the electromagnetic operator between kaon and pion states,''
  Phys.\ Rev.\ D {\bf 84}, 074503 (2011) 
  [arXiv:1108.1021 [hep-lat]].

\bibitem{kotonew}
N. Shimizu, ``Search for New Physics via the $K_L\to\pi^0\nu\bar\nu$ decay at the J-PARC KOTO
experiment'', talk at the 40th International Conference on High Energy Physics (ICHEP 2020),
Prague, Czech Republic, 30 July--5 August 2020.

\bibitem{Egana-Ugrinovic:2019wzj}
  D.~Egana-Ugrinovic, S.~Homiller, and P.~Meade,
``Light Scalars and the Koto Anomaly,''
  Phys.\ Rev.\ Lett.\  {\bf 124}, no. 19, 191801 (2020) 
  [arXiv:1911.10203 [hep-ph]].

\bibitem{Dev:2019hho}
  P.S.B.~Dev, R.N.~Mohapatra, and Y.~Zhang,
``Constraints on long-lived light scalars with flavor-changing couplings and the KOTO anomaly,''
  Phys.\ Rev.\ D {\bf 101}, no. 7, 075014 (2020) 
  [arXiv:1911.12334 [hep-ph]].

\bibitem{He:2020jly}
  X.G.~He, X.D.~Ma, J.~Tandean, and G.~Valencia,
``Evading the Grossman-Nir bound with $\Delta I=3/2$ new physics,''
  JHEP {\bf 2008}, 034 (2020) 
  [arXiv:2005.02942 [hep-ph]].

\bibitem{Littenberg:1995zy}
  L.S.~Littenberg and G.~Valencia,
``The decays $K\to\pi \pi \nu\bar\nu$ within the standard model'',
  Phys.\ Lett.\ B {\bf 385}, 379 (1996) 
  [hep-ph/9512413].

\bibitem{Chiang:2000bg}
  C.W.~Chiang and F.J.~Gilman,
``$K_{L,S}\to\pi\pi\nu\bar\nu$ decays within and beyond the standard model,''
  Phys.\ Rev.\ D {\bf 62}, 094026 (2000) 
  [hep-ph/0007063].

\bibitem{Adler:2000ic}
  S.~Adler {\it et al.} [E787 Collaboration],
``Search for the decay $K^+ \to \pi^+ \pi^0 \nu \bar{\nu}$,''
  Phys.\ Rev.\ D {\bf 63}, 032004 (2001) 
  [hep-ex/0009055].

\bibitem{E391a:2011aa}
  R.~Ogata {\it et al.} [E391a Collaboration],
``Study of the $K^0_L \to \pi^0 \pi^0 \nu \bar{\nu}$ decay,''
  Phys.\ Rev.\ D {\bf 84}, 052009 (2011) 
  [arXiv:1106.3404 [hep-ex]].

\bibitem{Gninenko:2014sxa}
  S.N.~Gninenko,
``Search for invisible decays of $\pi^0, \eta, \eta', K_S$ and $K_L$:
A probe of new physics and tests using the Bell-Steinberger relation,''
  Phys.\ Rev.\ D {\bf 91}, no. 1, 015004 (2015)  [arXiv:1409.2288 [hep-ph]].

\bibitem{Li:2016tlt}
  H.B.~Li,
``Prospects for rare and forbidden hyperon decays at BESIII'',
 Front.\ Phys.\ (Beijing) {\bf 12}, no. 5, 121301 (2017) 
  [arXiv:1612.01775 [hep-ex]]; (Erratum) {\bf 14}, 64001 (2019).

\bibitem{Ablikim:2019hff}
  M.~Ablikim {\it et al.},
``Future Physics Programme of BESIII'',
  Chin.\ Phys.\ C {\bf 44}, no. 4, 040001 (2020) 
  [arXiv:1912.05983 [hep-ex]].

\bibitem{Ambrosino:2019qvz}
  F.~Ambrosino {\it et al.} [KLEVER Project Collaboration],
``KLEVER: An experiment to measure BR($K_L\to\pi^0\nu\bar{\nu}$) at the CERN SPS,''
  arXiv:1901.03099 [hep-ex].

\bibitem{Luo:2018njj}
  Q.~Luo and D.~Xu,
``Progress on Preliminary Conceptual Study of HIEPA, a Super Tau-Charm Factory in China,''
in Proceedings of the 9th International Particle Accelerator Conference (IPAC2018),
Vancouver, Canada, 29 April--4 May 2018.

\bibitem{Barnyakov:2020vob}
  A.Y.~Barnyakov [Super Charm-Tau Factory Collaboration],
``The project of the Super Charm-Tau Factory in Novosibirsk,''
  J.\ Phys.\ Conf.\ Ser.\  {\bf 1561}, no. 1, 012004 (2020).

\end{thebibliography}
\end{document}